\documentclass[11pt]{article}
\pdfoutput=1
\usepackage[utf8]{inputenc}
\UseRawInputEncoding
\usepackage{verbatim,amsmath,amssymb}
\usepackage{epsfig,float,color}
\usepackage[font=small,labelfont=bf]{caption}
\usepackage{geometry}
\usepackage{setspace}
\usepackage{natbib}

\usepackage{amsthm,mathrsfs,amsfonts,dsfont} 
\usepackage{hyperref}
\usepackage{enumitem}
\usepackage[labelfont=bf]{caption}
\usepackage{subfig}
\usepackage{array}
\usepackage{makecell}
\usepackage{multirow}

\definecolor{darkblue}{rgb}{0,0,1}
\definecolor{col1}{rgb}{1,0,1}
\definecolor{col2}{rgb}{0,0.5,0}
\definecolor{col3}{rgb}{0.5,0,1}
\definecolor{col4}{rgb}{0.1,.75,0}

\hypersetup{pdftex=true, colorlinks=true, breaklinks=true, linkcolor=darkblue, menucolor=darkblue, pagecolor=darkblue, citecolor=darkblue, urlcolor=darkblue}

\newtheoremstyle{rem}
{6pt}
{6pt}
{\small}
{}
{\bf}
{:}
{.5em}
{}

\theoremstyle{rem}

\newcommand{\bitm}{\begin{itemize}}
\newcommand{\eitm}{\end{itemize}}
\newcommand{\bnumr}{\begin{enumerate}}
\newcommand{\enumr}{\end{enumerate}}

\newcommand {\eqb}[1]{\begin{equation}\begin{array}{#1}}
\newcommand {\eqe}{\end{array}\end{equation}}

\newcommand {\esb}[1]{\begin{equation*}\begin{array}{#1}}
\newcommand {\ese}{\end{array}\end{equation*}}
\newcommand {\ds}{\displaystyle}

\newcommand {\back}{\! \! \!}
\newcommand {\is}{\back &=& \back}

\newcommand {\II}{{I\kern-.3em I}}
\newcommand {\III}{{I\kern-.3em I\kern-.3em I}}

\newcommand {\mra}{\mathrm{a}}
\newcommand {\mrb}{\mathrm{b}}

\newcommand {\mre}{\mathrm{e}}

\newcommand {\IR}{{\rm\kern.24em
   \vrule width.02em height1.53ex depth-.05ex
   \kern-.3em R}}
\newcommand {\ic}{{\rm\kern.20em
   \vrule width.02em height1.0ex depth-.05ex
   \kern-.22em c}}
\newcommand {\ia}{{\rm\kern.20em
   \vrule width.02em height1.05ex depth-.0ex
   \kern-.25em a}}
\newcommand {\IC}{{\rm\kern.24em
   \vrule width.02em height1.4ex depth-.05ex
   \kern-.26em C}}
\newcommand {\ID}{{\rm\kern.34em
   \vrule width.02em height1.5ex depth-.05ex
   \kern-.36em D}}
\newcommand {\IS}{{\rm\kern.24em
   \vrule width.02em height1.6ex depth.05ex
   \kern-.26em S}}
\newcommand {\IT}{{\rm\kern.50em
   \vrule width.02em height1.55ex depth-.05ex
   \kern-.52em T}}

\newcommand {\IE}{{\rm\kern.24em
   \vrule width.02em height1.55ex depth-.05ex
   \kern-.33em E}}
\newcommand {\IEa}{{\rm\kern.24em
   \vrule width.02em height1.55ex depth-.05ex
   \kern-.33em E}^{1}_{ijkl}}
\newcommand {\IEb}{{\rm\kern.24em
   \vrule width.02em height1.55ex depth-.05ex
   \kern-.33em E}^{2}_{ijkl}}

\newcommand {\Ass}[2]{\kern 0.9ex \vrule width0.45em height0.2ex depth0ex \kern -2.1ex \bigwedge_{#1}^{#2}}
\newcommand {\ASS}[2]{\kern 1.45ex \vrule width0.5em height0.2ex depth0ex \kern -2.65ex \bigwedge_{#1}^{#2}}

\graphicspath{ {images/} }

\begin{document}

\begin{center}
\Large{\bf{Mechanical response of van der Waals and charge coupled carbon nanotubes}}\\

\end{center}

\renewcommand{\thefootnote}{\fnsymbol{footnote}}

\begin{center}
\large{Aningi Mokhalingam$^{\mra}$, Indranil S. Dalal$^{\mrb}$, and Shakti S.~Gupta$^{\mra}$\footnote[1]{corresponding author, email: ssgupta@iitk.ac.in}
}\\
\vspace{4mm}

\small{\textit{
$^\mra$Department of Mechanical Engineering, Indian Institute of Technology Kanpur, UP 208016, India \\[1.1mm]
$^\mrb$Department of Chemical Engineering, Indian Institute of Technology Kanpur, UP 208016, India \\[1.1mm]
}}

\end{center}

\vspace{-4mm}

\renewcommand{\thefootnote}{\arabic{footnote}}



\rule{\linewidth}{.15mm}
{\bf Abstract:}
This work investigates the mechanical response of single-walled carbon nanotubes (SWCNTs) coupled through van der Waals and electrostatic forces using molecular dynamic (MD) simulations and a continuum model. In MD simulations, the covalent bond interactions between the carbon atoms are modeled using ReaxFF potential. The dynamic charges, dependent on the local environment, are calculated employing the charge equilibrium formalism within the ReaxFF. In the continuum model, the SWCNTs are modeled using the geometrically nonlinear Euler-Bernoulli beam theory. The Galerkin's approach is used to discretize the equations of motion. An approximate model to account for the end charge concentration in the SWCNTs, calibrated from the MD data, is incorporated into the beam model. The pair of SWCNTs are prescribed with two sets of boundary conditions: Fixed-fixed and fixed-free. The pull-in voltages at which the two SWCNTs snap onto each other with fixed-fixed boundary conditions obtained from the MD simulations agree within an error of $\sim 0.5 \%$ with that computed from the nonlinear beam theory. For fixed-free boundary conditions, the role of geometric nonlinearity is found to be insignificant. However, for this case, the concentrated charges play a significant role in determining the pull-in voltages. The post-pull-in response of the SWCNTs for both boundary conditions is investigated in detail through the MD simulations. The post-pull-in results presented here can be used as a benchmark for results obtained from continuum models in the future. Further, the proposed research helps design nano-resonators/tweezers/switches.

{\bf Keywords:} Carbon nanotube; Euler-Bernoulli beam theory; molecular dynamics; pull-in voltage; resonator; tweezer.

\vspace{-5mm}
\rule{\linewidth}{.15mm}

\vspace{-2mm}

\section{Introduction}\label{s:intro}
Single-walled carbon nanotubes (SWCNTs) have potential applications in electronic devices and developing nanoelectromechanical systems (NEMS) such as resonators, switches, tweezers, sensors, memory devices, etc \citep{Cumings2000, Kis2006, Kim1999}. The SWCNT-based resonators and switches may consist of two or more slender elements with either \textit{fixed-fixed} or \textit{fixed-free} boundary conditions. On the other hand, the SWCNT-based tweezers comprise of two elements having \textit{fixed-free} boundary conditions. Micro- and nano-resonators find applications in sensing mass \citep{Chiu2008, Mate2005}, strain \citep{Woj2004, Beeby2000}, and pressure or temperature \citep{Welham1996, Thundat1994}. In single-element resonators, a shift in the resonant frequency is measured to identify the perturbations in the system. In the multiple-element resonators, coupled either electrically or mechanically, changes in the response amplitudes are used to assess the perturbations. In the latter case of resonators, the shift provided by the response amplitudes over the baseline signature could be three times higher than the shift in the resonant frequency \citep{Thiru2009}, thereby increasing its sensitivity. The nano-tweezers used to manipulate matter at a nano length scale \citep{Kim1999} are very similar to the coupled resonators. They consist of two elements separated at a distance. Externally applied voltage between them leads to their deflection towards each other, which creates sufficient grip to hold the matter of interest. 

The knowledge of pull-in voltage, that is the critical voltage at which the gap between two elements collapses suddenly, is essential for the safe operation of resonators and ensuring a firm grip in nano-tweezers. This sudden collapse of gap is termed as \textit{pull-in instability} in the literature.

The pull-in instabilities of coupled-resonators and tweezers have been studied experimentally  \, \citep{Kim1999, Akita2002, Ke2005, Chang2009}, through continuum models \citep{Wang2004, Ramezani2011, Farrokhabadi2013, Farrokhabadi2014, Keivani2017, Zare2017, Yekrangisendi2019, Bianchi2020}, and by performing molecular simulations \citep{Aluru2004, FAKHRABADI201338, HWANG2005163, sircar2020}. 

\citet{Kim1999} fabricated a multiwalled carbon nanotube (MWCNT) tweezer and demonstrated that it is capable of manipulating submicron clusters and nanowires. \citet{Akita2002} developed a nano-tweezer consisting of MWCNTs attached to a Si tip with metal electrodes. They reported that the manipulation performance of the tweezer strongly depends on the substrate and the target material's van der Waal (vdW) attraction forces. \citet{Chang2009} fabricated tweezers using nanowires and provided a relation between the applied electrostatic force and the corresponding deflection of the nanowires.

\citet{Ramezani2011} studied the pull-in instabilities of beam-plate type nano-tweezers using distributed and lumped models and reported that the lumped model underestimates the \textit{pull-in voltages} than those determined from the distributed models. Using a beam model, \citet{Wang2004} investigated the pull-in instability of tweezer-like nanostructures under the influence of vdW interactions. They have employed Galerkin's method to discretize the coupled differential equations and determined the critical length for specified radii of tubes at which pull-in occurs. \citet{Farrokhabadi2014} have incorporated the electrostatic forces along with the vdW attraction forces in their continuum model to analyze the static response of SWCNT tweezer. Using the Euler--Bernoulli beam theory (EBBT) hypothesis, they studied the static and dynamic pull-in behavior and reported that the tweezing range depends upon the gap between the two SWCNTs. \citet{Zare2017} reported the influence of vdW attraction forces and surface ripples formed due to bending of the SWCNTs on the \textit{pull-in voltage} using differential quadrature method. \citet{Ke2005a} proposed an approximate model to determine the end concentration charge at the tip of a freely-standing SWCNT above an infinite grounded plane using classical electrostatics. In their another work, \citet{Ke2005b} investigated the effect of charge concentration on the \textit{pull-in voltage} of a cantilever SWCNT switch.  

\citet{Aluru2004} studied the static and dynamic pull-in instability of SWCNT-based switch using molecular dynamic (MD) simulations and the EBBT. In MD simulations, they used continuum formulations to prescribe the electrostatic and vdW forces. \citet{FAKHRABADI201338} studied the influence of defects in SWCNT on the pull-in behavior employing AIREBO potential. The simulation set-up in this study comprised of static and oppositely charged SWCNT-substrate pair. Using Tersoff potential, \citet{HWANG2005163}, \citet{sircar2020} studied the pull-in instabilities of a NEM switch assuming a constant potential along the SWCNT.  Based on this assumption \citet{HWANG2005163} applied a constant mechanical load and \citet{sircar2020} used static charge to mimic the electrostatic force.

The mechanical deformations of SWCNTs under the influence of an externally applied electric field with applications related to resonators, tweezers, and switches have been reported widely in the literature. In most of the cases, the SWCNTs are modeled as beams, and the focus is up to the pull-in voltage. Moreover, the charges on the SWCNTs and hence on beams are taken to be static. However, towards developing a rigorous model for a complete range of operations until the failure of a device, the dynamic charge distribution due to the deforming shape of SWCNT until pull-in voltage and circumferential and the neutral axis deformations post-pull-in voltage must be studied. In this work, we use MD simulations employing the ReaxFF potential and EBBT to study the complete range of operations of SWCNT-based resonators and tweezers and explore what happens if the limits on applied voltages are exceeded. Two sets of boundary conditions are considered: Fixed-fixed and fixed-free, for resonator and tweezers, respectively. The MD simulations compute environment-dependent dynamic charges. The equations of motion obtained through EBBT are solved using Galerkin's method. Use of conventional eigenfunctions in Galerkin's method is found to show numerical instabilities, which are overcome by employing modified eigenfunctions for slender beams.

The organization of this paper is as follows: Section~\ref{Molecular simulations} describes the molecular simulations, setup, and charge calculations. A brief introduction to the continuum model is given in Section~\ref{Continuum model}. The numerical results are discussed in Section~\ref{Results-p2}, followed by conclusions in Section~\ref{Conclusions}.
\section{Molecular simulations} \label{Molecular simulations}
In MD simulations, the atomic interactions between the carbon atoms are defined using ReaxFF forcefield \citep{Chenoweth2008, Strachan2003}. The total energy of a system $E_{\text{system}}$ is given by
\eqb{lll}
\ds E_{\text{system}}=E_{\text{bond}}+E_{\text{over}}+E_{\text{under}}+E_{\text{val}}+E_{\text{tors}}+E_{\text{vdW}}+E_{\text{coul}}+E_{\text{Specific}}~,
\eqe
where $E_{\text{bond}}, E_{\text{over}}, E_{\text{under}}, E_{\text{val}}, E_{\text{tors}}, E_{\text{vdW}}, E_{\text{coul}}$ are the energies associated with the bond stretch, over- and under-coordination, angle-bending, torsional-bending, van der Waals and Coulomb interactions, respectively. $E_{\text{specific}}$ consists of hydrogen binding, lone pair, conjugations and $\text{C}_2$ corrections, needed for the stability of specific systems. The ReaxFF potential uses the instantaneous atomic pair distance to calculate the bond order $BO_{ij}'$ at each time step. The bond order includes the contributions from the $\sigma$, $\pi$ and $\pi \pi$ bonds and is given as
\eqb{lll} \label{bond_order}
\displaystyle BO_{ij}' & =BO_{ij}^{\sigma}+BO_{ij}^{\pi}+BO_{ij}^{\pi \pi} \\
& =\exp\left[p_{bo1} \left(\frac{r_{ij}}{r_o^{\sigma}}\right)^{p_{bo2}}\right] 
+\exp\left[p_{bo3} \left(\frac{r_{ij}}{r_o^{\pi}}\right)^{p_{bo4}}\right]+\exp\left[p_{bo5} \left(\frac{r_{ij}}{r_o^{\pi \pi}}\right)^{p_{bo6}}\right]~, 
\eqe
where $r_{ij}$ is the interatomic distance between atoms $i$ and $j$, $r_o$ is the equilibrium distance between them, and $p_{bo}$ terms are potential constants. The over- and under-coordination terms in Eq.~\eqref{bond_order} impose penalty energies to the system to enforce the proper number of bonds. The nonbonded interactions- van der Waals and Coulomb, calculate the energy between all pairs of atoms, including bonded ones. To avoid the excess repulsive energy between the paired atoms, the nonbonded interaction terms are shielded by introducing an extra variable (see  \citep{Chenoweth2008, Strachan2003} for details).

The atomic charges ($q_i$), dependent on the local environment, are calculated using QEq formalism \citep{Rappe1991}. The total potential energy of a system can be written as $E_{\text{tot}}=E_0(x)+E_{\text{es}}(q,x)$, where $E_0(x)$ is the sum of nonionic terms such as energies associated with the bond stretching, angle bending, torsional bending, etc., and $E_{\text{es}}(q,x)$ is the electrostatic energy. The term $E_{\text{es}}(q,x)$ is dependent on atomic charge ($q_i|_{i=1,2,.., N}$) and their positions (${\bf{x}}_i|_{i=1,2,.., N}$), where N is the number of atoms and is given as
\eqb{lll} \label{eq:elec}   
\ds E_{\text{es}} \is \ds \sum_i \chi_i q_i+\sum_i \frac{1}{2}\eta_iq_i^2 +\sum_{(i,j)} J_{ij} q_iq_j~,
\eqe
where $\chi_i$ and $\eta_i$ are electronegativity and self-Coulomb potential of atom $i$, respectively, and $J$ depends on the position of atoms $i$ and $j$. The first two terms in Eq.~\eqref{eq:elec} denote the intra-atomic electrostatic energies and the last term Coulomb interaction energy. Atomic chemical potential ($\mu$) is then obtained by differentiating Eq.~\eqref{eq:elec} with respect to $q_i$, as
\eqb{lll} \label{eq:ene_2}
\ds \mu_i \is \ds \sum_i \chi_i +\sum_i \eta_iq_i +\sum_{(i,j)} {J}_{ij} q_j~,
\eqe
Following the condition of equalization of the atomic chemical potential at the equilibrium leads to 
\eqb{lll}  \label{eq:ns}
\ds \mu_1=\mu_2=....=\mu_N
\eqe
The above condition gives $N-1$ equations. Now, applying the charge neutrality constraint, i.e.,
\eqb{lll}  \label{eq:neu}
\ds \sum_i^N q_i=0
\eqe
At the beginning of each timestep, Eqs.~\eqref{eq:ns} and \eqref{eq:neu}, $N$ linear equations in $q_i$, are solved to calculate the charge on each atom for a given structure. 
\subsection{MD simulation setup}
The coupled resonator/tweezer is modeled using two parallel (7,7) SWCNTs with length to diameter ratio $\sim$ 15 to enable the Euler-Bernoulli beam theory (EBBT) applicable for developing an equivalent continuum model. 
The following steps are followed in the MD simulations: 
\begin{enumerate}[label=(\roman*)]
\item individual SWCNTs are relaxed separately. This is achieved by using the Polak-Ribiere’s conjugate gradient energy minimization method \citep{Polak1969} within the energy and force tolerances of $10^{-10}$ and $10^{-10}$ Kcal/mole/\AA, respectively, followed by thermalization at 0.1 K using Nos\'e-Hoover \citep{Evans1985} thermostat with three Nos\'e-Hoover chains.
\item the two relaxed SWCNTs are arranged in the vertical plane as shown in Fig.~\ref{fig:res1}. The fixed end boundary conditions are applied by constraining all degrees of freedom of the edge atoms. 
\begin{figure}[!htbp]
\centering
\subfloat[\label{f_1a}]{%
  \includegraphics[trim=4 0 00 00,clip,height=35mm]{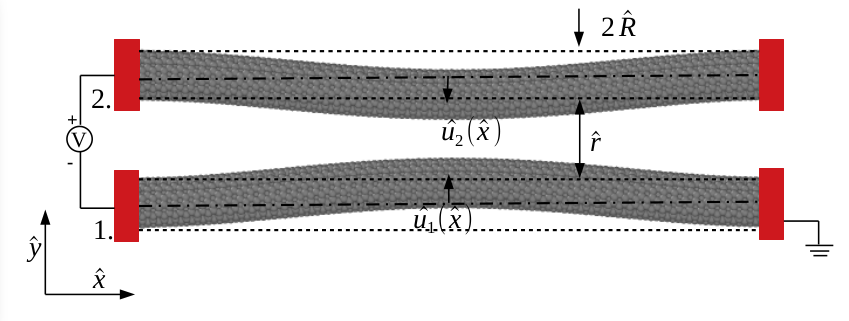}%
}\hfill
\subfloat[\label{f_1b}]{%
  \includegraphics[trim=1 0 00 00,clip,height=35mm]{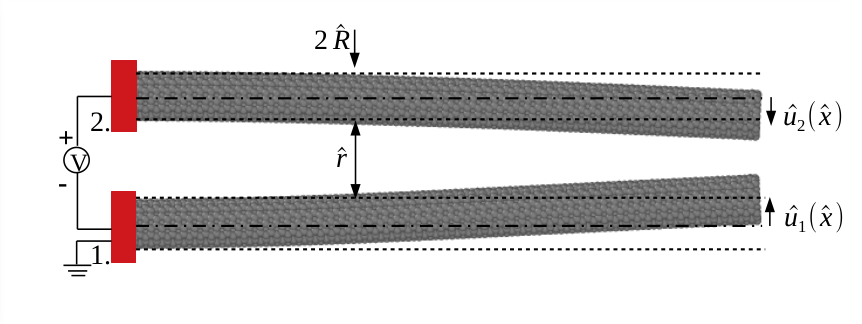}%
}\hfill 
     \caption{Schematic representation of a coupled-resonator made of (7,7) single-walled carbon nanotubes with (a) fixed-fixed and (b) fixed-free boundary conditions. The red colored blocks providing fixed boundary conditions are conducting. \label{fig:res1}}
\end{figure}
\item the charge distribution on each SWCNT due to external potential $V$ is obtained by providing an offset electronegativity \citep{Chen2009charge, Onofrio2015}. 
\item after calculating the charge on each atom using QEq formalism at every timestep, the system is brought to equilibrium configuration at 0.1 K using the Nos\'e-Hoover thermostat. 
\end{enumerate}
All the simulations are performed using an open-source code, LAMMPS \citep{lammps}.
\subsection{Computation of charge distribution}
\begin{figure}[!htbp]
\centering
  \includegraphics[width=0.5\columnwidth]{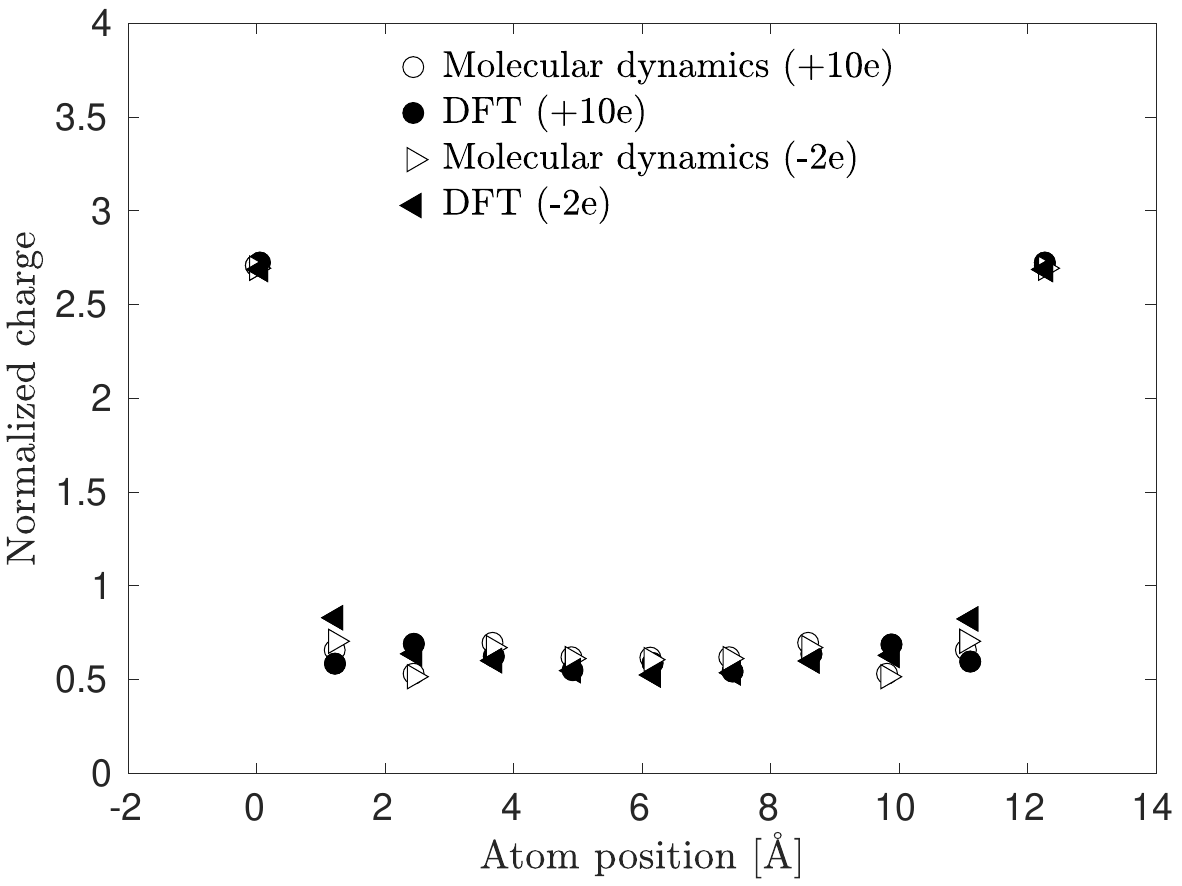}%
\caption{Comparision of charge distribution determined from the MD simulations and DFT \citep{Keb2002} for 1.2 nm long (5,5) SWCNT.    \label{dft_charge}}
\end{figure}

In MD simulations, the two SWCNTs are assigned with the electronegativities of $\chi_0+V/2$ and $\chi_0-V/2$, where $\chi_0=5.7254$ eV \citep{Strachan2003}, the standard potential parameter for Carbon is used and $V$ is varied as a parameter. For reasonably accurate charge distribution along SWCNTs, $85\%$ of the SWCNT's length is chosen as the cutoff distance in the simulations. The charge distribution obtained in our MD simulations for a free-standing (5,5) SWCNT above an infinite grounded plane compares well (within $\approx 3 \%$) with that determined from DFT simulations \citep{Keb2002}, as shown in Fig.~\ref{dft_charge}, thus validating the simulation procedure. 

In the beam model, we assume the SWCNTs to be an axially connected array of unit-length conducting rigid rings. The capacitance of two parallel unit length rings is given by \citep{Hayt_1981}
\eqb{lll}
\ds C(\hat{r}) = \frac{2 \pi \epsilon_0}{\ln\left[1+\frac{\hat{r}}{\hat{R}}+\sqrt{\left(\frac{\hat{r}}{\hat{R}}+1\right)^2-1}\right]}~,
\label{e:capacitance_i}
\eqe
where $\epsilon_0=8.854 \times 10^{-12} \, \text{C}^2\text{N}^{-1}\text{m}^{-2}$, the permitivity of vaccum. It is noted here that $\hat{R}$ and $\hat{r}$ are the radii of the SWCNTs and surface-to-surface distance between them, respectively (see Fig.~\ref{fig:res1}). The average charge per atom $Q_{\text{atom}}^c$ is then calculated as
\eqb{lll}
\ds Q_{\textnormal{atom}}^c=\frac{C(\hat{r})V\hat{L}}{N_t}~,
\label{e:ch_at}
\eqe
where $V$ is the applied voltage, $\hat{L}$ is length of the SWCNT and $N_t$ is total number of atoms. In the MD simulations, the average charge per atom ${Q}_{atom}^m$ at a given axial location is calculated as
\eqb{lll}
\ds \bar{Q}_{\textnormal{atom}}^m=\frac{{\sum_{i=1}^{N_r} Q_i^m}}{N_r} ~,
\eqe
where $N_r$ is the number of ring/circumferential atoms with charge $Q_{i}^m$ ($i=1$ to $N_r$) at the given axial location.

\begin{figure}[h]
        \centering
    \includegraphics[height=60mm]{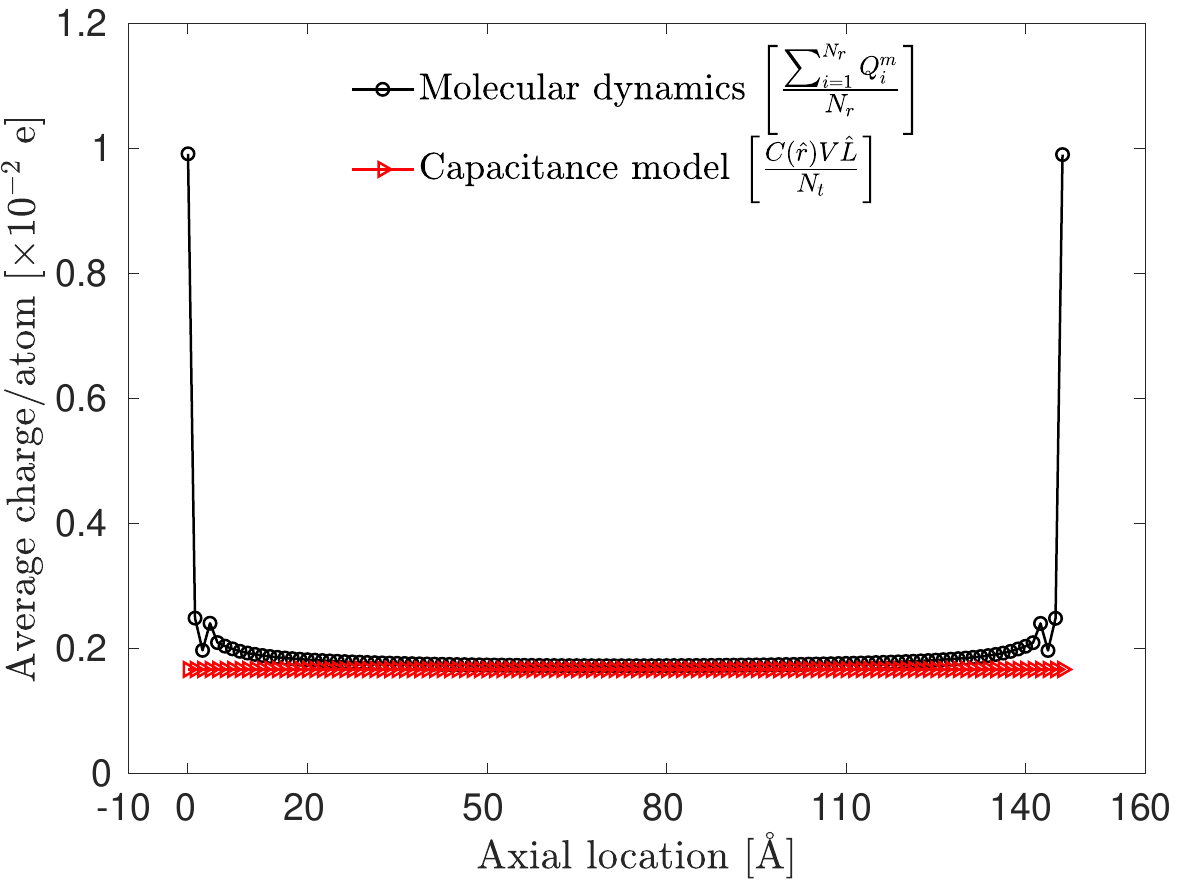}
    \vspace{-3mm}
    \caption{Comparison of average charge/atom computed employing Eq.~\eqref{e:capacitance_i} and MD simulations for (7,7) SWCNT. The tube is placed 1 nm above the half-space (symmetric mid-plane for the two SWCNTs) and is subject to a potential difference of +1 V. In MD simulations average charge/atom is computed considering the charges $Q^m$ on the number of atoms $N_r$ in the ring/circumference of the tube at a given axial location. In the capacitance model the total number of atoms $N_t$ is taken from the atomic structure used for MD simulations.\label{fig:charge}}
\end{figure}
As seen in Fig.~\ref{fig:charge}, the charge calculated from the capacitance model \eqref{e:ch_at} agrees well with that from MD simulations, except at the ends. We fix this mismatch by scaling $Q_{\textnormal{atom}}^c$ appropriately at the ends. Towards this, we have carried out MD simulations on SWCNTs of radii between 0.34 to 3.4 nm and the gap between them ranging from 1 nm to 5 nm. From the obtained data, the scaling factor $f$ in terms of normalized quantities, $R'$ and $ r'$, is found to be 
\eqb{lll}
\displaystyle 
\displaystyle  f(\hat{R}, \hat{r})= c_1 + c_2 R' + c_3 r' + c_4 R'r'~,
\label{e:charge_end_atom}
\eqe
where $c_I=$ 3.88, 0.38, 0.26 and 0.16 for $I=1$, 2, 3, and 4 are constants, $ R'=\frac{\hat{R}}{ \hat{R}(\text{SWCNT}(7,7))},$ and $ r'=2\hat{r}(\text{in nm})/\hat{g}_0$, where $\hat{g}_0$ is the initial separation distance in our study i.e., 2 nm. Now, the total charge accumulated over the SWCNT is given by
\eqb{lll}
\displaystyle 
\ds Q^{Total} \is \ds Q_{\textnormal{atom}}^c\,(N_t-2\,N_r^\mre) + Q^{\mre}\,2\,N_r^{\mre}\,,
\label{e:charge_end1}
\eqe
where $N_r^\mre$ is the number of end atoms and $Q^{\mre}$ is the average charge/atom concentrated at the end, which is given by
\eqb{lll}
\displaystyle 
\ds Q^\mre \is \ds \,f(\hat{R},\hat{r})\,\frac{C(\hat{r})\,V\,\hat{L}}{N_t}\,,
\label{e:charge_end12}
\eqe  
From Eq.~\eqref{e:charge_end1} and \eqref{e:charge_end12}, the capacitance/length with end concentration charge is given by 
\eqb{lll}
\displaystyle 
\displaystyle C'(\hat{r}) = C(\hat{r})\left[\frac{N_t-2N_r^e}{N_t}+\frac{f(\hat{R},\hat{r})\,\hat{L}\,N_r^e}{N_t}\left(\delta(\hat{x}-\hat{x}_0)+\delta(\hat{x}-\hat{x}_L)\right)\right]~,
\label{e:charge_total}
\eqe
where $\hat{x}_0$ and $\hat{x}_L$ are the end locations of an SWCNT. In the next section while developing an equivalent continuum model for the SWCNTs if their ends are clamped the end charge correction term will not be applied. However, if the ends are free, Eq.~\eqref{e:charge_total} will be employed. 
\section{Beam model} \label{Continuum model}
The SWCNTs are modeled using Euler-Bernoulli beam theory (EBBT). The coupled equations of motion of these SWCNTs or beams of length $\hat{L}$, cross-section area $\hat{A}$, are given as \citep{fakhrabadi2014non, Ouakad2009, Sedighi}
\eqb{lll} \label{e:eq_motion}
\displaystyle \hat{E}\hat{I}\frac{\partial^4 \hat{u}_J}{\partial{\hat{x}^4}}+\hat{c}\frac{\partial \hat{u}_J}{\partial{\hat{t}}}+\hat{\rho}\hat{A}\frac{\partial^2 \hat{u}_J}{\partial{\hat{t}^2}}-\hat{F}_{non}= -(-1)^J\hat{F}_{tot}~,
\eqe 
where ${\hat{u}_J}$ ($J=1$ and 2) are the transverse deflections (along $\hat{y}$-direction, see Fig.~\ref{fig:res1}) of the two SWCNTs. Further, ${\hat{E}}$, ${\hat{I}}$, $\hat{\rho}$, and $\hat{c}$ are Young's modulus, second moment of area, mass density, and damping coefficient per unit length, respectively. The  nonlinear terms, $\hat{F}_{non}$, for the two boundary conditions are given as 
\begin{equation}
\displaystyle \hat{F}_{non}=\left[\frac{\hat{E}\hat{A}}{2L} \int_0^{\hat{L}} \left(\frac{\partial \hat{u}_{J}}{\partial{\hat{x}}} \right)^2 d\hat{x}\right]\frac{\partial^2 \hat{u}_{J}}{\partial{\hat{x}^2}}~,
\end{equation}
for fixed-fixed boundary conditions, and 
\eqb{lll}
\displaystyle \hat{F}_{non}= -EI \frac{\partial}{\partial \hat{x}} \left[\frac{\partial \hat{u}_J}{\partial \hat{x}} \frac{\partial}{\partial \hat{x}}\left( \frac{\partial \hat{u}_J}{\partial \hat{x}} \frac{\partial^2 \hat{u}_J}{\partial \hat{x}^2}\right) \right]~,
\eqe
for fixed-free boundary conditions. The total non-contact force/length, $\hat{F}_{tot}$, is summation of the electrostatic force, $\hat{F}_{\text{elec}}, $ and van der Waal force, $\hat{F}_{\text{vdW}}$. The electrostatic force $\hat{F}_{\text{elec}}$, between the two SWCNTs is calculated through the capacitance model. For an applied voltage $V$, the electrostatic energy is given as ${E_{\text{elec}}^{\text{fixed-fixed}}}=\frac{1}{2}C(\hat{r})V^2$ and ${E_{\text{elec}}^{\text{fixed-free}}}=\frac{1}{2}C'(\hat{r})V^2$  for fixed-fixed and fixed-free boundary conditions, respectively. It should be noted that only the fixed-free boundary condition would require the end charge correction. The electrostatic force/length is then computed as follows \citep{Farrokhabadi2014} 
\eqb{lll}
\begin{split}
\displaystyle \hat{F}_{\text{elec}}= & -\frac{d( E_{elec})}{d\hat{r}} \\
 = & \frac{\pi \epsilon_0 V^2 K }{\sqrt{(\hat{r}+\hat{R}-\hat{u}_1+\hat{u}_2)^2-R^2} \left[\ln\left(1+\frac{\hat{r}-\hat{u}_1+\hat{u}_2}{\hat{R}}+\sqrt{\left(1+\frac{\hat{r}-\hat{u}_1+\hat{u}_2}{\hat{R}} \right)^2-1 }\right)\right]^2}
~, 
\end{split}
\label{e:ele_force}
\eqe
where $K=1$ for fix-fixed and $(\frac{N_t-N_r^e}{N_t}+\frac{\hat{L}fN_r^e}{N_t}\delta(\hat{x}-\hat{L}))$ for fixed-free boundary conditions, respectively. The coarse-grained (CG) 18-24LJ potential is used to define the van der Waal interactions between the two SWCNTs. The van der Waals energy/length is given by \citep{ji2019}
\eqb{lll}
\ds \hat{E}_{\text{vdW}}=\frac{256}{27}\epsilon_{18-24}\hat{\rho}_1^2\left(\frac{88179 \pi\sigma^{24}_{18-24}}{2^{19}(\hat{r}+2\hat{R}-\hat{u}_1+\hat{u}_2)^{23}}-\frac{6435 \pi\sigma^{18}_{18-24}}{2^{15}(\hat{r}+2\hat{R}-\hat{u}_1+\hat{u}_2)^{17}} \right)~,
\eqe
where $\hat{\rho_1}$ is the line density, determined as $ {1}/{b_0}$, with $b_0$ as 5 \AA \ \citep{ji2019}. The LJ parameters $\epsilon_{18-24}$ and  $\sigma_{18-24} $ are, respectively, 10.7 Kcal/mole and 1.23 nm, which are calibrated directly from the MD data (see Appendix~\ref{appendix_b-1} for the details). $\hat{F}_{\text{vdW}}$ is calculated as follows 
\eqb{lll}
\begin{split}
\displaystyle \hat{F}_{\text{vdW}} =  & -\frac{d(\hat{E}_{\text{vdW}})}{d\hat{r}}  \\
= &  \frac{256}{27}\epsilon_{18-24}\hat{\rho}_1^2\left(\frac{88179 \times 23 \pi\sigma^{24}_{18-24}}{2^{19}(\hat{r}+2\hat{R}-\hat{u}_1+\hat{u}_2)^{24}}-\frac{6435 \times 17 \pi\sigma^{18}_{18-24}}{2^{15}(\hat{r}+2\hat{R}-\hat{u}_1+\hat{u}_2)^{18}} \right)~,
\label{e:vdW_force}
\end{split}
\eqe
Now, the boundary conditions for the fixed-fixed beam \citep{fakhrabadi2014non} are 
\eqb{lll}
\begin{split}
\ds \hat{u}(0,\hat{t})&=\hat{u}(\hat{L},\hat{t})=0; \, \frac{\partial {\hat{u}(0,\hat{t})}}{\partial {\hat{x}}}=\frac{\partial {\hat{u}(\hat{L},\hat{t})}}{\partial {\hat{x}}}=0~,
\end{split}
\label{e:ff_bcs}
\eqe
and for the fixed-free beam \citep{fakhrabadi2014non} are
\begin{equation}
\begin{split}
\ds \hat{u}(0,\hat{t})=0; \, \frac{\partial {\hat{u}(0,\hat{t})}}{\partial {\hat{x}}} =0; \, \frac{\partial^2 {\hat{u}(\hat{L},\hat{t})}}{\partial {\hat{x}^2}} =0; \,\frac{\partial^3 {\hat{u}(\hat{L},\hat{t})}}{\partial {\hat{x}^3}}=0~,
\end{split}
\label{e:c_bcs}
\end{equation}
The following non-dimensional quantities are considered for convenience: 
\eqb{lll}
\begin{split}
\displaystyle u=\frac{\hat{u}}{\hat{r}} ~, \, x = \frac{\hat{x}}{\hat{L}}~, \, t =  \frac{\hat{t}}{\hat{T}}~, 
\end{split}
\label{e:non_dim}
\eqe
where $\hat{T}=\sqrt{{\hat{\rho}\hat{A}\hat{L^4}}/{\hat{E}\hat{I}}}$ is a time constant. The non-dimensional equations of motion are then given by
\eqb{lll} \label{non_eom}
 \ds u_J''''+c\dot{u}_J+\ddot{u}_J-F_{non}=-(-1)^J F_{tot}~, \eqe
where
\eqb{lll} \label{non}
\ds F_{non}=\left[\lambda_1 \int_0^1 (u_J')^2dx \right]u_J''~, 
\eqe 
for fixed-fixed boundary conditions, and
\eqb{lll}
\ds F_{non}= \lambda_4^2 \left[u_J'\left(u_J'u_J'' \right)'\right]'~,
\eqe
for fixed-free boundary conditions. The dot and prime over the quantities represent the partial derivatives with respect to time and space, respectively. The non-dimensional electrostatic and van der Waal forces are  
\eqb{lll}
\displaystyle {F}_{\text{elec}}=\frac{\lambda_2 K}{\sqrt{(1+R_0-u_1+u_2)^2-R_0^2} \left[\ln\left(1+\frac{1-u_1+u_2}{R_0}+\sqrt{\left(1+\frac{1-u_1+u_2}{R_0} \right)^2-1 }\right)\right]^2}~,
\label{e:mod_ele_force}
\eqe 
and
\eqb{lll}
\displaystyle {F}_{\text{vdW}} = \lambda_3\left(\frac{88179 \times 23 \sigma^{24}_{0}}{2^{19}(1+2R_0-{u}_1+{u}_2)^{24}}-\frac{6435 \times 17 \sigma^{18}_{0}}{2^{15}(1+2{R}_0-{u}_1+{u}_2)^{18}} \right) ~.
\label{e:mod_vdW_force}
\eqe
In Eqs.~(\ref{non} - \ref{e:mod_vdW_force}), 
$
\ds R_0 = \frac{\hat{R}}{\hat{r}}~, c=\frac{\hat{c}\hat{L}^4}{\hat{E}\hat{I}\hat{T}}~, \, \lambda_1 = \frac{\hat{A}\hat{E}\hat{r}^2}{2\hat{E}\hat{I}}~, \,
\lambda_2 = \frac{\pi \epsilon_0 V^2 \hat{L}^4}{2\hat{r}^2\hat{E}\hat{I}}~, 
\lambda_3=\frac{256 \pi \hat{L}^4\epsilon_{18-24} \hat{\rho_1}^2 }{27\hat{r}\hat{E}\hat{I}}~,  \lambda_4=\frac{\hat{r}}{\hat{L}}~, \,$ and $ \ds \sigma_0=\frac{\sigma_{18-24}}{\hat{r}}~.
$

\subsection{Solution procedure}
The governing equations of motion are discretized using Galerkin's method. Towards that, the solution of Eq.~\eqref{non_eom} is approximated as 
\eqb{lll}\label{e:dyn:sol1}
\ds u_1(x,t)=\ds \sum_{i=1}^m q_i(t)\phi_i(x)~,
\eqe
and
\eqb{lll}\label{e:dyn:sol2}
\ds u_2(x,t)=\ds \sum_{j=1}^m p_j(t)\phi_j(x)~,
\eqe
where $m$ is the mode number, $p_i$ and $q_j$ are the time varying non-dimensional modal coordinates, and $\phi_i$ and $\phi_j$ are the basis functions, which are discussed subsequently. The solution is then obtained by substituting Eq.~\eqref{e:dyn:sol1} and Eq.~\eqref{e:dyn:sol2} into Eq.~\eqref{non_eom} and multiplying with $\phi_n$ and thereafter integrating within the boundary limits. The resulting equations for the fixed-fixed boundary condition are:  
\eqb{lll} 
\begin{split}
\ds \omega_n^2 q_n +c\dot{q}_n+\ddot{q_n}  = &  \ds \lambda_1 q_iq_jq_k \int_0^1 \phi_i' \phi_j'dx \int_0^1\phi_k{''}\phi_ndx \\
& \ds + \int_0^1 \frac{\lambda_2 K}{\sqrt{(1+R_0- q_i(t)\phi_i(x)+ p_j(t)\phi_j(x))^2-R_0^2}} \\
& \ds \times \frac{1}{\left[\ln\left(1+\frac{1- q_i(t)\phi_i(x)+ p_j(t)\phi_j(x)}{R_0}+\sqrt{\left(1+\frac{1- q_i(t)\phi_i(x)+ p_j(t)\phi_j(x)}{R_0} \right)^2-1 }\right)\right]^2}\phi_ndx \\
& \ds +\int_0^1 \lambda_3\left(\frac{88179 \times 23 \sigma^{24}_{0}}{2^{19}(1+2R_0-q_i(t)\phi_i(x)+ p_j(t)\phi_j(x))^{24}}
 \right. \\
& \ds \left. -\frac{6435 \times 17 \sigma^{18}_{0}}{2^{15}(1+2{R}_0-q_i(t)\phi_i(x)+ p_j(t)\phi_j(x))^{18}} \right)\phi_ndx~, 
\end{split} \label{eq:dyn_ff1}
\eqe
\eqb{lll} 
\begin{split}
\ds \omega_l^2 p_l+c\dot{p}_l+\ddot{p_l}  = & \ds \lambda_1 p_ip_jp_k \int_0^1 \phi_i' \phi_j'dx \int_0^1\phi_k{''}\phi_ldx \\
& \ds -\int_0^1 \frac{\lambda_2 K}{\sqrt{(1+R_0- q_i(t)\phi_i(x)+ p_j(t)\phi_j(x))^2-R_0^2}} \\
& \ds \times \frac{1}{\left[\ln\left(1+\frac{1- q_i(t)\phi_i(x)+ p_j(t)\phi_j(x)}{R_0}+\sqrt{\left(1+\frac{1- q_i(t)\phi_i(x)+ p_j(t)\phi_j(x)}{R_0} \right)^2-1 }\right)\right]^2}\phi_ndx \\
& \ds -\int_0^1 \lambda_3\left(\frac{88179 \times 23 \sigma^{24}_{0}}{2^{19}(1+2R_0-q_i(t)\phi_i(x)+ p_j(t)\phi_j(x))^{24}}
 \right. \\
& \ds \left. -\frac{6435 \times 17 \sigma^{18}_{0}}{2^{15}(1+2{R}_0-q_i(t)\phi_i(x)+ p_j(t)\phi_j(x))^{18}} \right)\phi_ndx~, 
\end{split} \label{eq:dyn_ff2}
\eqe
and those for the fixed-free boundary condition are: 
\eqb{lll}
\begin{split}
\ds \omega_n^2 q_n+c\dot{q}_n +\ddot{q_n}= & \ds \lambda_4^2q_iq_jq_k \int_0^1 [\phi_i' (\phi_j'\phi_k{''})']'\phi_ndx \\
 & \ds   +\int_0^1 \frac{\lambda_2 K}{\sqrt{(1+R_0- q_i(t)\phi_i(x)+ p_j(t)\phi_j(x))^2-R_0^2}} \\
 & \ds \times \frac{1}{\left[\ln\left(1+\frac{1- q_i(t)\phi_i(x)+ p_j(t)\phi_j(x)}{R_0}+\sqrt{\left(1+\frac{1- q_i(t)\phi_i(x)+ p_j(t)\phi_j(x)}{R_0} \right)^2-1 }\right)\right]^2}\phi_ndx \\
 & \ds +\int_0^1 \lambda_3\left(\frac{88179 \times 23 \sigma^{24}_{0}}{2^{19}(1+2R_0-q_i(t)\phi_i(x)+ p_j(t)\phi_j(x))^{24}}
 \right. \\
 & \ds \left. -\frac{6435 \times 17 \sigma^{18}_{0}}{2^{15}(1+2{R}_0-q_i(t)\phi_i(x)+ p_j(t)\phi_j(x))^{18}} \right)\phi_ndx~,  
\end{split} \label{eq:dyn_cc1}
\eqe
 \begin{equation}\label{eq:dyn_cc2}
 \begin{split}
\ds \omega_l^2 p_l+c\dot{p}_l+\ddot{p_l} =  & \ds \lambda_4^2p_ip_jp_k  \int_0^1 [\phi_i' (\phi_j'\phi_k{''})']'\phi_ldx \\
 & \ds - +\int_0^1 \frac{\lambda_2 K}{\sqrt{(1+R_0- q_i(t)\phi_i(x)+ p_j(t)\phi_j(x))^2-R_0^2}} \\
 & \ds \times \frac{1}{\left[\ln\left(1+\frac{1- q_i(t)\phi_i(x)+ p_j(t)\phi_j(x)}{R_0}+\sqrt{\left(1+\frac{1- q_i(t)\phi_i(x)+ p_j(t)\phi_j(x)}{R_0} \right)^2-1 }\right)\right]^2}\phi_ndx \\
 & \ds -\int_0^1 \lambda_3\left(\frac{88179 \times 23 \sigma^{24}_{0}}{2^{19}(1+2R_0-q_i(t)\phi_i(x)+ p_j(t)\phi_j(x))^{24}}
 \right. \\
 & \ds \left. -\frac{6435 \times 17 \sigma^{18}_{0}}{2^{15}(1+2{R}_0-q_i(t)\phi_i(x)+ p_j(t)\phi_j(x))^{18}} \right)\phi_ndx~, 
 \end{split}
\end{equation}
The above equations are numerically solved using ODE15s integrator in MATLAB \citep{matlab}.
\subsection{Mode shapes and their numerical stability} \label{a_b_modes}
In this work, linear, undamped modes shapes, obtained by solving the undamped homogeneous part of the governing Eq.~\eqref{non_eom}, are used in the Galerkin's method. The closed form of mode shapes for the fixed-fixed boundary conditions is \citep{rao2019vibration}:
\begin{equation}\label{eb:mode_ff}
\begin{split}
\ds \phi_m(x)&=\ds \sinh(\alpha_m x)-\sin(\alpha_m x) \\
&-\left[\frac{\sin(\alpha_m)-\sinh(\alpha_m)}{\cos(\alpha_m)-\cosh(\alpha_m)}\right]\cosh(\alpha_m x)-\cos(\alpha_m x)~,
\end{split}
\end{equation}
where $\alpha_m$ are the roots following the characteristic equation:
\begin{equation}
\ds \cos{(\alpha_m)}\, \cosh{(\alpha_m)}-1=0~,
\end{equation}
and for the fixed-free boundary conditions \citep{rao2019vibration}:
\begin{equation}\label{eb:mode_ffree}
\begin{split}
\ds \phi_m(x)&=\ds \sinh(\tilde{\alpha}_m x)-\sin(\tilde{\alpha}_m x) \\
&-\left[\frac{\sinh(\tilde{\alpha}_m)+\sin(\tilde{\alpha}_m)}{\cosh(\tilde{\alpha}_m)+\cos(\tilde{\alpha}_m)}\right]\cosh(\tilde{\alpha}_m x)-\cos(\tilde{\alpha}_m x)~,
\end{split}
\end{equation}
where $\tilde{\alpha}_m$ are the roots following characteristic equation:
\begin{equation}
\ds \cos(\tilde{\alpha}_m)\, \cosh(\tilde{\alpha}_m)+1=0~,
\end{equation}
\begin{figure}[!htbp]
\centering
  \includegraphics[width=0.7\columnwidth]{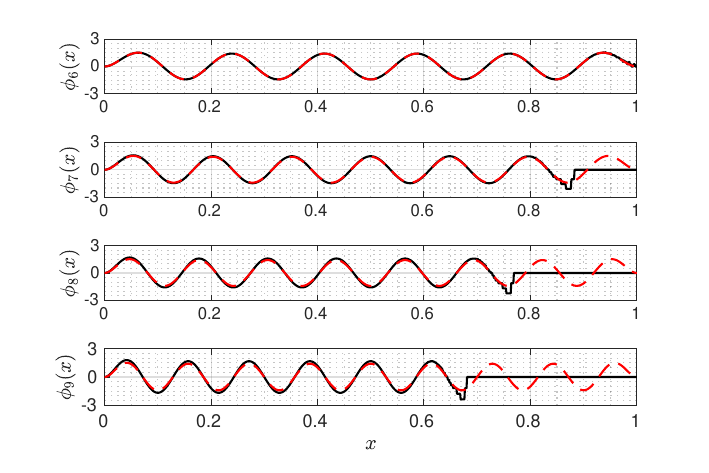}%
     \caption{Comparison between higher symmetric modes for fixed-fixed boundary condition obtained from Eqs.~\eqref{eb:mode_ff} (black continuous line) and \eqref{e:mode_ff_mod} (red dashed line). \label{fig:modes_gan}}
\end{figure}
When two identical fixed-fixed SWCNTs are used in the resonator, they deform symmetrically with respect to their geometric center because the electrostatic load is symmetric. Therefore, studying their behavior using an equivalent continuum beam model will require summing over only the symmetric modes in Eqs.~\eqref{e:dyn:sol1} and \eqref{e:dyn:sol2}. In order to obtain converged results, several higher symmetric modes will be needed. But, the modes given by Eq.~\eqref{eb:mode_ff} suffer from numerical issues with increasing values of $\alpha_m$ \citep{GONCALVES2007}. However, when using identical fixed-free SWCNTs in the resonator, there is no restriction on using a particular mode type employing the equivalent beam model. To circumvent the numerical issues embedded in Eq.~\eqref{eb:mode_ff} at higher modes, that is, for higher values of $\alpha_m$, we will use the modes proposed by \citet{GONCALVES2007}. These are given as follows: 
\begin{equation}\label{e:mode_ff_mod}
\begin{split}
\ds \phi_m(x)&=\ds \exp(-\beta_m x)-\cos(\beta_m x) 
+\left(1+\frac{\exp(-\beta_m)-\cos(\beta_m)+\sin(\beta_m}{\sinh(\beta_m)-\sin(\beta_m)} \right)\sin(\beta_m x) \\
&-\left(\exp(-\beta_m)-\cos(\beta_m)+\sin(\beta_m) \right)\exp(\beta_m(x-1))~,
\end{split}
\end{equation}
where $\beta_m$ are the roots following characteristic equation:
\begin{equation}
    \ds \cos(\beta_m)\,\cosh(\beta_m)-1=0~,
\end{equation}
The symmetric modes obtained from Eqs.~\eqref{eb:mode_ff} and \eqref{e:mode_ff_mod} are compared in Fig.~\ref{fig:modes_gan} to demonstrate numerical issues in the classical beam modes at higher $\alpha_m$. However, the lower modes given in Eq.~\eqref{e:mode_ff_mod} have the error of order $10^{-2}$ \citep{GONCALVES2007}. Therefore, in Eqs.~\eqref{e:dyn:sol1} and \eqref{e:dyn:sol2}, we use the first, third, fifth, seventh, and ninth symmetric modes computed from Eq.~\eqref{eb:mode_ff}, and if required (as in case of convergence studies), we use the additional symmetric modes obtained from Eq.~\eqref{e:mode_ff_mod}.
\subsection{Computation of material properties for the beam model}
We employ mechanics of solids based approach to compute $\hat{E}\hat{I}$ and $\hat{A}\hat{E}$, the bending and axial stiffnesses, respectively, used in Eq.~\eqref{non_eom}. The value of $\hat{E}\hat{I}$ is determined by first computing the maximum transverse deflection of the fixed-fixed SWCNT using MD to the applied uniformly distributed load ($w_u)$ of 0.0157 N/m and then using it in the beam deflection equation ($\delta_{max}={w_u\hat{L}^4}/{384\hat{E}\hat{I}}$) \citep{archer2012}. Similarly, the value of $\hat{A}\hat{E}$ is determined by first computing the axial deflection of the fixed-free SWCNT using MD to the applied axial load of 20 nN and then using it in the bar elongation formula (see Eq.~2.2, Pg.~83 of \citet{archer2012}). The values of $\hat{E}\hat{I}$ and $\hat{A}\hat{E}$ for (7,7) SWCNT are found to be $1.15\times 10^{-25} \ \textnormal{N-m}^2$ and $0.91 \times 10^{-8}$ N, respectively.
\begin{figure}[!htbp]
\centering
  \includegraphics[width=0.5\columnwidth]{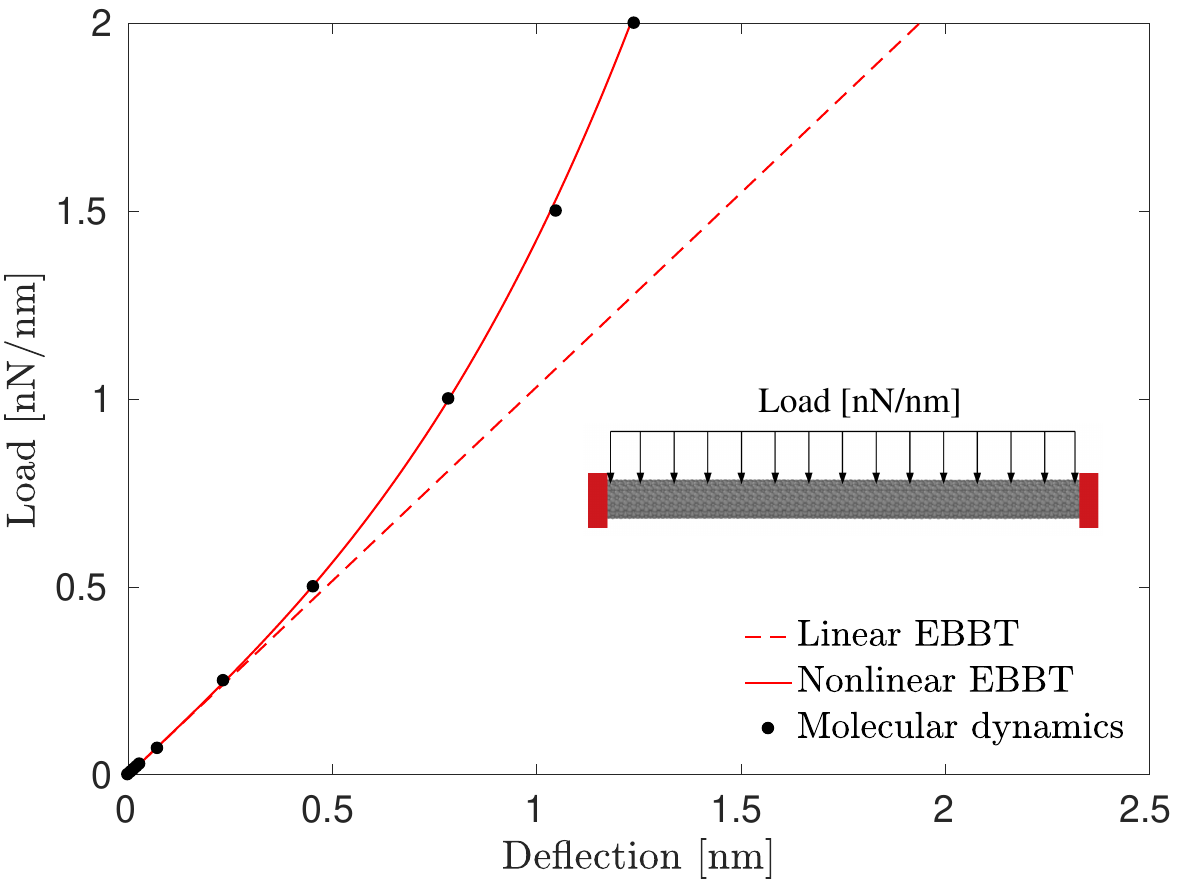}%
\caption{Deflection at mid-span of the SWCNT under uniformly distributed load    \label{prop_cnt}}
\end{figure}
Further, to check the beam model's applicability at higher deflections, we incrementally increased the vertical uniformly distributed load up to 2 nN/nm. The maximum deflections (at mid-span) obtained from the MD simulations and the EBBT, using $\hat{E}\hat{I}$ and $\hat{A}\hat{E}$, found above, are shown in Fig.~\ref{prop_cnt}. As seen, the MD simulation results agree with the linear EBBT only up to $\approx 0.35$ nm mid-span deflection and deviate after that. In contrast, the nonlinear EBBT captures the nonlinearity sufficiently well. 
\section{Numerical results} \label{Results-p2}
In this section, we present pull-in and post pull-in behaviors of the SWCNTs determined from MD and EBBT. The charge distribution in SWCNTs during the process of deformation is obtained entirely through MD. The validity and limitations of the beam model in pre- and post-pull-in phases are also discussed.
\subsection{Fixed-fixed boundary conditions}
The deflections of the two SWCNTs and their beam model are studied up to the \textit{pull-in voltage}, at which the two SWCNTs/beams come closest to each other, starting from the gap between them, $\hat{g}_0 =$ 2 nm. In MD simulations, the voltage is applied gradually with an increment of 2 V for the resonator with fixed-free boundary conditions. At each applied voltage, the two SWCNTs are allowed to settle to their equilibrium configurations, and their atomic positions are used to determine the deflections. In EBBT, Eqs.~\eqref{eq:dyn_ff1} and \eqref{eq:dyn_ff2} are solved numerically.

\subsubsection{Pull-in analysis} 
\begin{figure}[!htbp]
\centering
\subfloat[]{%
  \includegraphics[width=0.48\columnwidth]{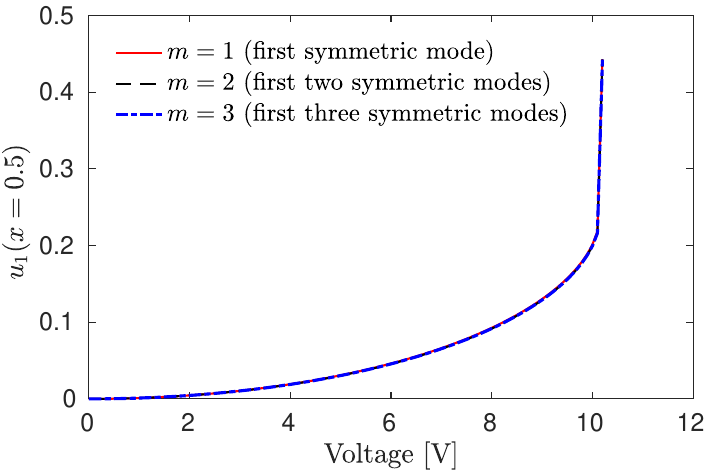}%
}\hfill
\subfloat[]{%
  \includegraphics[width=0.48\columnwidth]{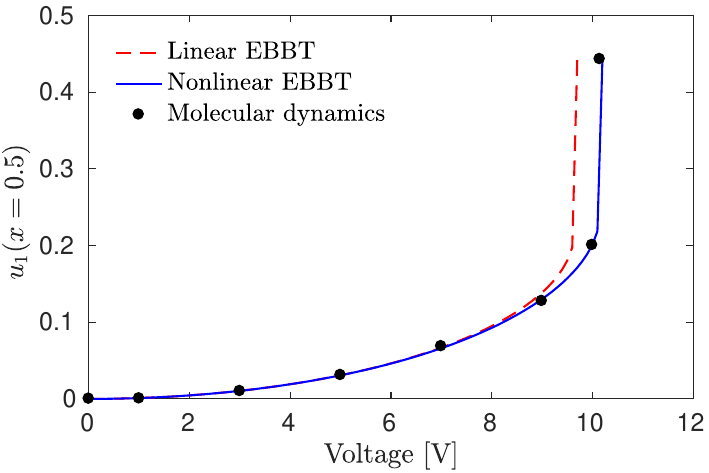}%
}\hfill 
     \caption{Mid-span deflection of fixed-fixed SWCNT-1 with increasing voltage. (a)~The mid-span deflection obtained using various combinations of symmetric modes in the solution of nonlinear EBBT. (b)~Comparison of the mid-span deflection versus voltage, from MD, linear and nonlinear EBBT. \label{fig:disp}}
\end{figure}

In the beam model, we have used only symmetric modes due to symmetric loading and symmetric boundary conditions. To check the convergence, we first calculate the mid-span deflection employing first one, two, and three symmetric modes. This is shown in Fig.~\ref{fig:disp}(a). This confirms that one mode approximation provides the converged results, agreeing with \citet{Ouakad2009}.

In Fig.~\ref{fig:disp}(b), we show the comparison of mid-span deflections of SWCNT-1 with the increasing applied voltage. The values of \textit{pull-in voltage} determined from the linear and nonlinear EBBT and MD simulations are found to be 9.7 V, 10.2 V, and 10.15 V, respectively. The value of \textit{pull-in voltage} and deflection at mid-span obtained from MD simulation agree well with those determined from the nonlinear EBBT within an average error of $0.5\%$ and $3\%$, respectively. As seen from Fig.~\ref{fig:disp}(b), the results for the linear and nonlinear beam models agree only up to the non-dimensional deflection of 0.06 due to the absence of stretching non-linearity in the linear theory. Therefore, for predicting the correct value of \textit{pull-in voltage} for a resonator with a large initial gap, the usage of nonlinear beam model is imperative. 

\subsubsection{Post-pull-in analysis}
\begin{figure}[!htbp]
\centering
\subfloat[]{%
  \includegraphics[width=0.48\columnwidth]{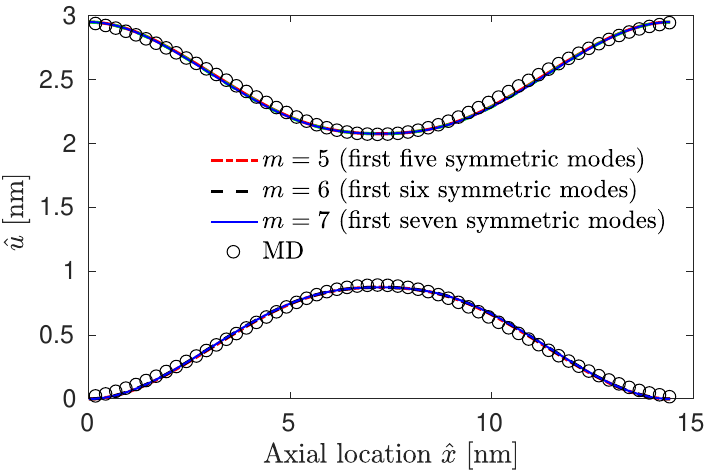}%
}\hfill
\subfloat[]{%
  \includegraphics[width=0.48\columnwidth]{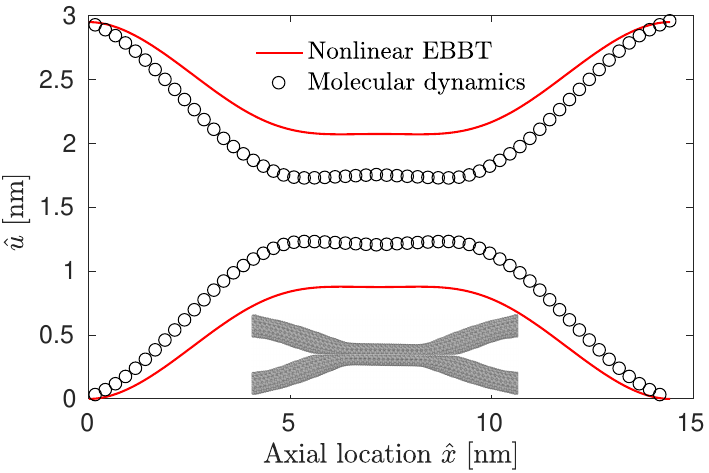}%
}\hfill 
     \caption{Deflections of beam-1 and -2 for fixed-fixed boundary conditions: (a)~using various combinations of symmetric modes in the solution of nonlinear EBBT and MD at \textit{pull-in voltage} ($\approx 10.2$ V) and (b)~at $V=15$ V. (Equilibrium configuration obtained from MD is shown in in-set). \label{fixed_Contact}}
\end{figure}
Figure~\ref{fixed_Contact}(a) shows deformation of the neutral axis determined from the nonlinear EBBT model employing the first five, six, and seven symmetric modes and that obtained from MD simulations at \textit{pull-in voltage}. As discussed earlier, one mode approximation predicts the converged \textit{pull-in voltage} and the mid-span deflections. However, for an applied voltage corresponding to pull-in and beyond, at least seven symmetric modes are required in the beam model (see Fig.~\ref{fixed_Contact}(a)). The mid-section of contacting SWCNTs becomes flat when the applied voltage is more than the pull-in voltage. To predict this behavior, the beam model requires more and more symmetric modes in the solution. The contact lengths and the transverse deflections of two SWCNTs determined from the nonlinear EBBT and MD are found to differ by $\sim 5\%$ and $ \sim 0.6\%$, respectively, as shown in Fig.~\ref{fixed_Contact}(a). To investigate the reasons for these differences, we recorded the time history of the transverse and circumferential deformations of the SWCNTs, and the interaction forces between them, from the MD simulations. 
\begin{figure}[ht]
\centering
  \includegraphics[width=0.5\columnwidth]{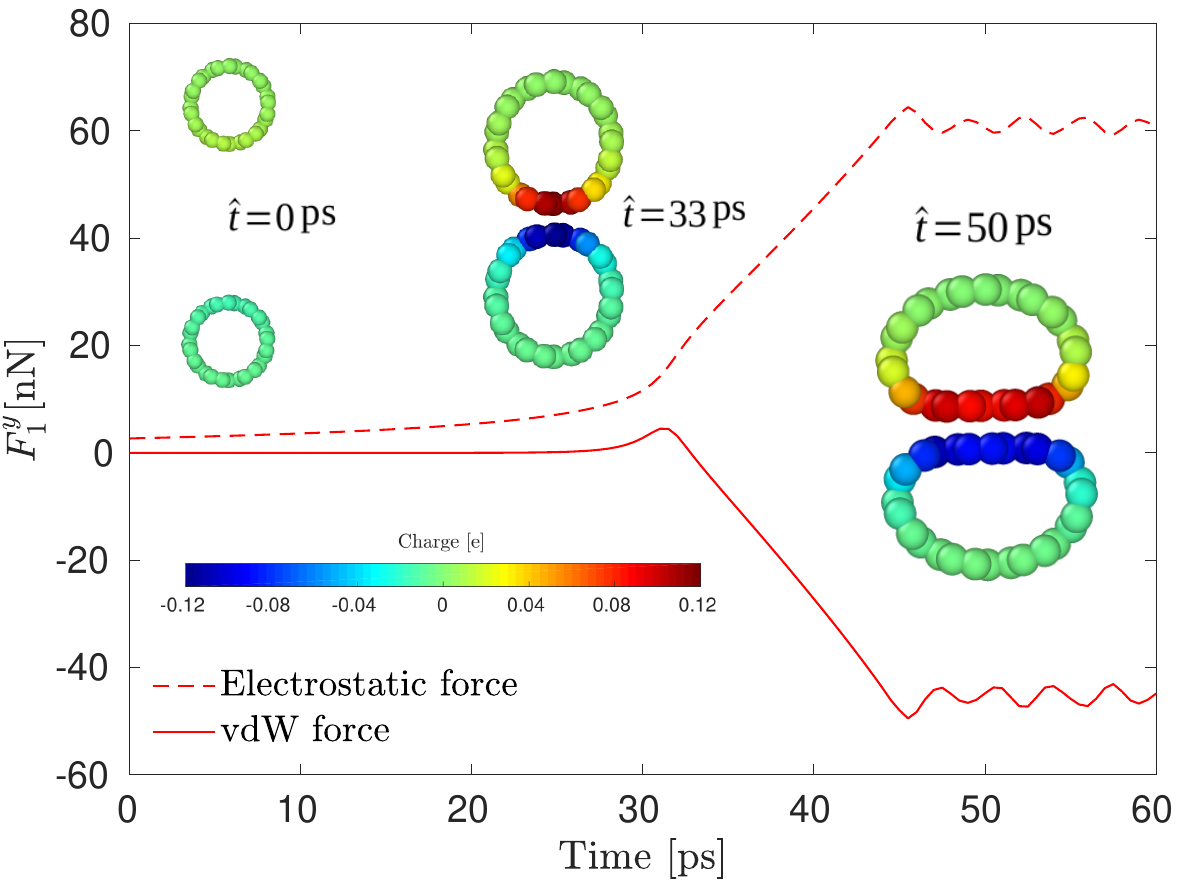}%
\caption{Fixed-fixed SWCNTs: Time variation of the electrostatic and vdW forces on the SWCNT-1 when the applied voltage equals the \textit{pull-in voltage}. Also shown are the cross-section at the mid-span of the SWCNTs at different times.     \label{f_e_v}}
\end{figure}
In Fig.~\ref{f_e_v}, we have shown the cross-sections of the SWCNTs at their mid-span and charge distribution at three different time instants using a color map. It is evident that the vdW and electrostatic forces rise gradually as the separation distance between the two SWCNTs decreases. At the pull-in, the drastic rise in the electrostatic forces led the two SWCNTs to deflect beyond the vdW equilibrium distance starting from the mid-span, which generates the vdW repulsive forces, as shown in Fig.~\ref{f_e_v}. These oppositely concentrated forces at the mid-span of the SWCNTs distort the cross-section to non-circular (see inset of Fig.~\ref{f_e_v}). The resulting deformed SWCNTs induce higher vdW and electrostatic forces than those in the beam model, where the cross-section does not deform by virtue of the assumptions in the kinematics. The deviations become more for the higher applied voltages than the \textit{pull-in voltage}. Figure~\ref{fixed_Contact}(b) compares the neutral axis deformation determined using nonlinear EBBT employing the first seven symmetric modes and MD simulations for an applied voltage equal to 15 V, higher than that at pull-in. The contact lengths determined from both methods are found to differ by $\sim 10.4\%$ due to significant deformations in the cross-section of the SWCNTs (see inset of Fig.~\ref{fixed_Contact}(b)), making their centroidal lines to move towards each other. 

Next, only using MD simulations, we study the variation of charge distribution with time at the \textit{pull-in voltage}. 
\begin{figure}[h]
\centering
\subfloat[]{%
  \includegraphics[width=0.49\columnwidth]{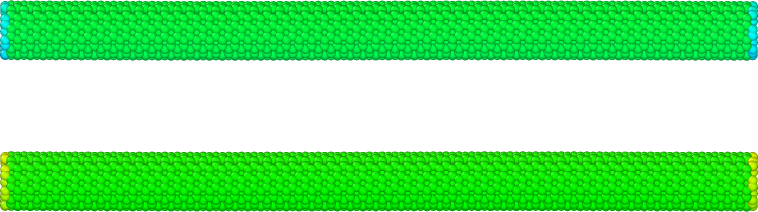}%
}\hfill
\centering
\subfloat[]{%
  \includegraphics[width=0.49\columnwidth]{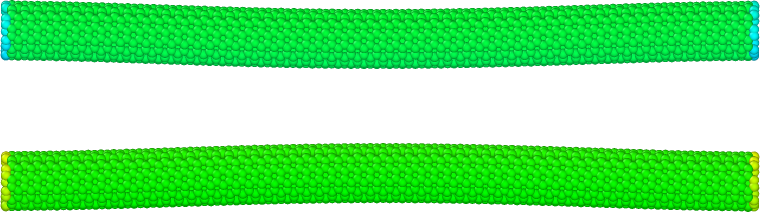}%
}\hfill
\centering
\subfloat[]{%
  \includegraphics[width=0.49\columnwidth]{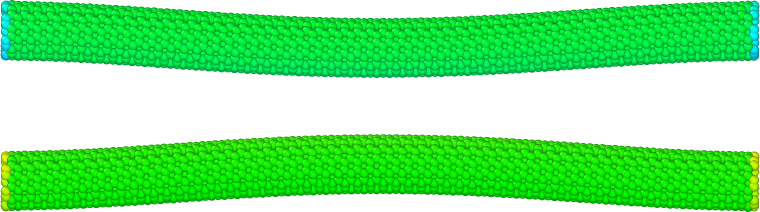}%
}\hfill
\centering
\subfloat[]{%
  \includegraphics[width=0.49\columnwidth]{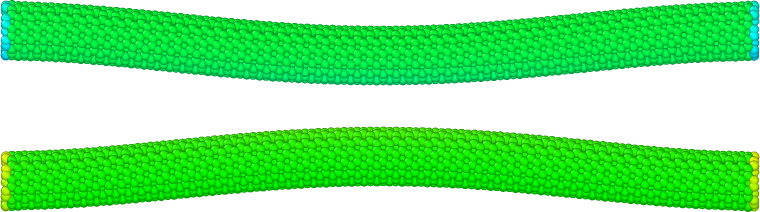}%
}\hfill
\centering
\subfloat[]{%
  \includegraphics[width=0.49\columnwidth]{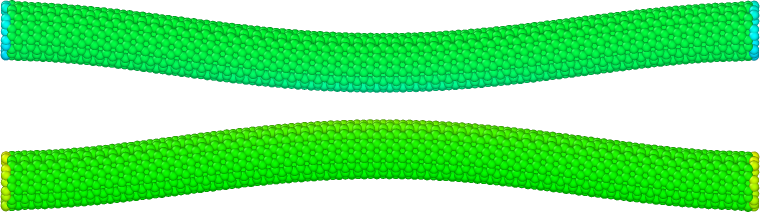}%
}\hfill
\centering
\subfloat[]{%
  \includegraphics[width=0.49\columnwidth]{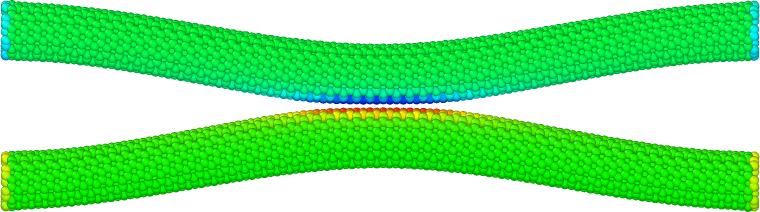}%
}\hfill
\centering
\subfloat{%
  \includegraphics[width=0.7\columnwidth]{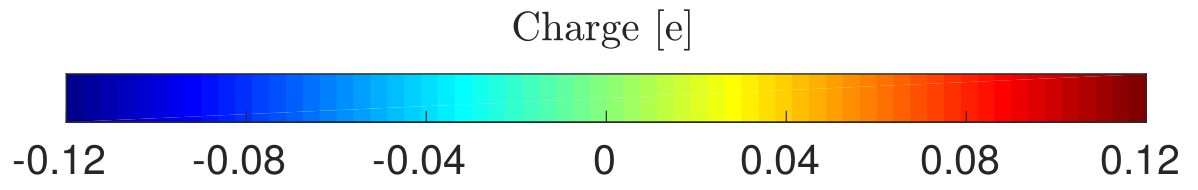}%
}\hfill
     \caption{Variation in configuration and charge contour with time (a) 0 ps (b) 7 ps (c) 14 ps (d) 21 ps (e) 28 ps and (f) 33 ps at \textit{pull-in voltage}.\label{fig:config_f20}}
\end{figure}
At $\hat{t}=0$ ps, when both SWCNTs are parallel and straight, the charge gets  distributed uniformly along the length, except at the ends, as shown in Fig.~\ref{fig:config_f20}(a). However, as the two SWCNTs deflect towards each other, we notice high charge concentration near the mid-span (see Fig.~\ref{fig:config_f20}). This is due to the continuous growth of electrostatic energy ($E_i^e(q,\hat{r})$) with decreasing of $\hat{r}$. At the pull-in state, the atoms at the mid-span, which are in contact with the second SWCNT, charge $150 \%$ more than the end atoms (see Fig.~\ref{fig:config_f20}).
\subsubsection{Limitations in the higher applied voltages}
The end concentration charge governs the structural instabilities of free-standing SWCNTs to the applied external voltage. The SWCNT leads to structural damage by ejecting the atoms from the tips when the applied voltage exceeds the critical value \citep{Keb2002, wang2001}. 
\begin{figure}[!htbp]
\begin{center} \unitlength1cm
\begin{picture}(0,5)
\put(-4,1.6){\includegraphics[height=27.5mm]{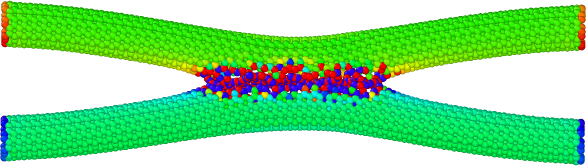}}
\put(-4,0){\includegraphics[height=15mm]{f_bar.png}}
\end{picture}
\caption{Configuration of SWCNTs when the applied voltage is equal to 20 V.} 
\label{f_20_config}
\end{center}
\end{figure}
However, in deformed SWCNTs with fixed-fixed boundary conditions, the structural instabilities are governed by the mid-span atoms, particularly those coming in contact with the second SWCNT. For an applied voltage of 20 V for the geometry of SWCNTs and $\hat{g}_0$ considered here, the induced electrostatic forces at the mid-span exceed the binding energy of carbon atoms, causing structural damage in the SWCNTs, as shown in Fig.~\ref{f_20_config}. 

\subsection{Fixed-free boundary conditions}
This section describes the pre- and post pull-in behavior of the tweezer with fixed-free boundary conditions. The transverse separation gap between two SWCNTs is 2 nm. In MD simulations, the voltage is applied gradually with an increment of 0.2 V, whereas in EBBT, Eqs.~\eqref{eq:dyn_cc1} and \eqref{eq:dyn_cc2} are solved numerically.
\subsubsection{Pull-in analysis}
\begin{figure}[!htbp]
        \centering
 \includegraphics[height=60mm]{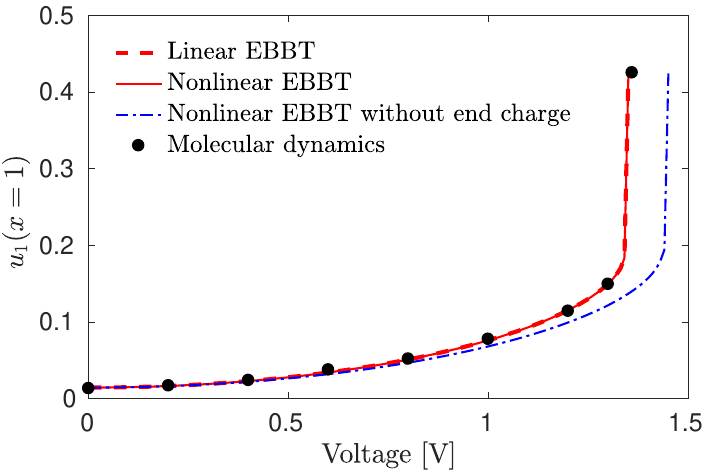}
         \vspace{-3mm}
\caption{Variation of tip deflection ($u_1$) with the applied voltage. \label{charge_effect}}
\end{figure}

Figure~\ref{charge_effect} shows variation of the free end deflection with the applied voltage determined from the linear EBBT with end charge concentration, nonlinear EBBT with and without end charge concentration, and MD simulations.
One mode approximation is used in the EBBT models. The pull-in voltages determined from the nonlinear EBBT without end concentration charge and MD are 1.52 and 1.4 V, respectively, which differ $\sim 8.5\%$. The approximate end charge concentration model reduces this error to $\sim 1\%$, as shown in Fig.~\ref{charge_effect}. Thus, the inclusion of the end charge concentration plays a significant role in predicting the correct pull-in voltages. Further, the nonlinearity has shown no effect on the \textit{pull-in voltage} for the studied $\hat{g_0}$. 

\subsubsection{Post-pull-in analysis}
\begin{figure}[ht]
\centering
\subfloat[]{%
  \includegraphics[width=0.48\columnwidth]{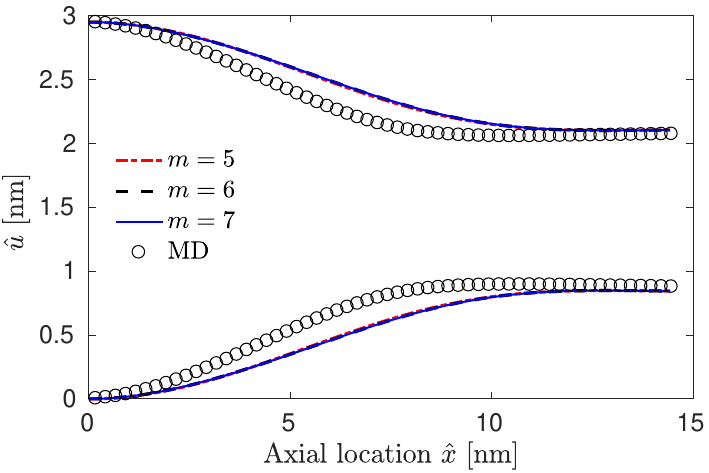}%
}\hfill
\subfloat[]{%
  \includegraphics[width=0.48\columnwidth]{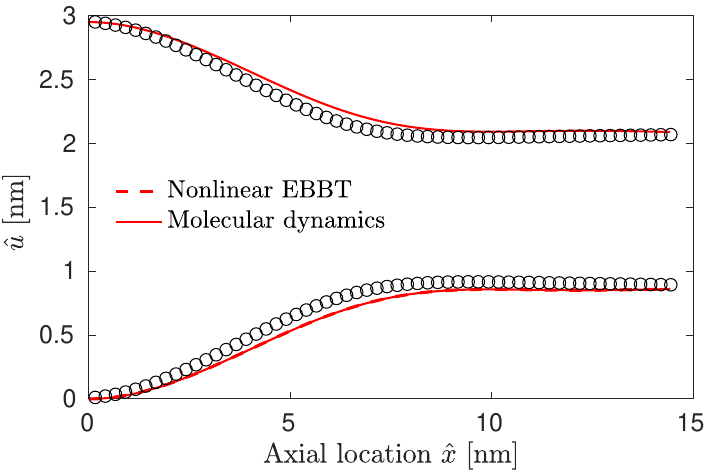}%
}\hfill 
     \caption{Deflections of beam-1 and -2 for fixed-free boundary conditions: (a) using 5-mode,
6-mode, and 7-mode approximations in the solution of nonlinear EBBT and MD at \textit{pull-in voltage} ($\sim 1.4$) V and (b) at $V=5$ V. \label{fixed_free_Contact}}
\end{figure}

\begin{figure}[!htbp]
\centering
\subfloat[]{%
  \includegraphics[width=0.48\columnwidth]{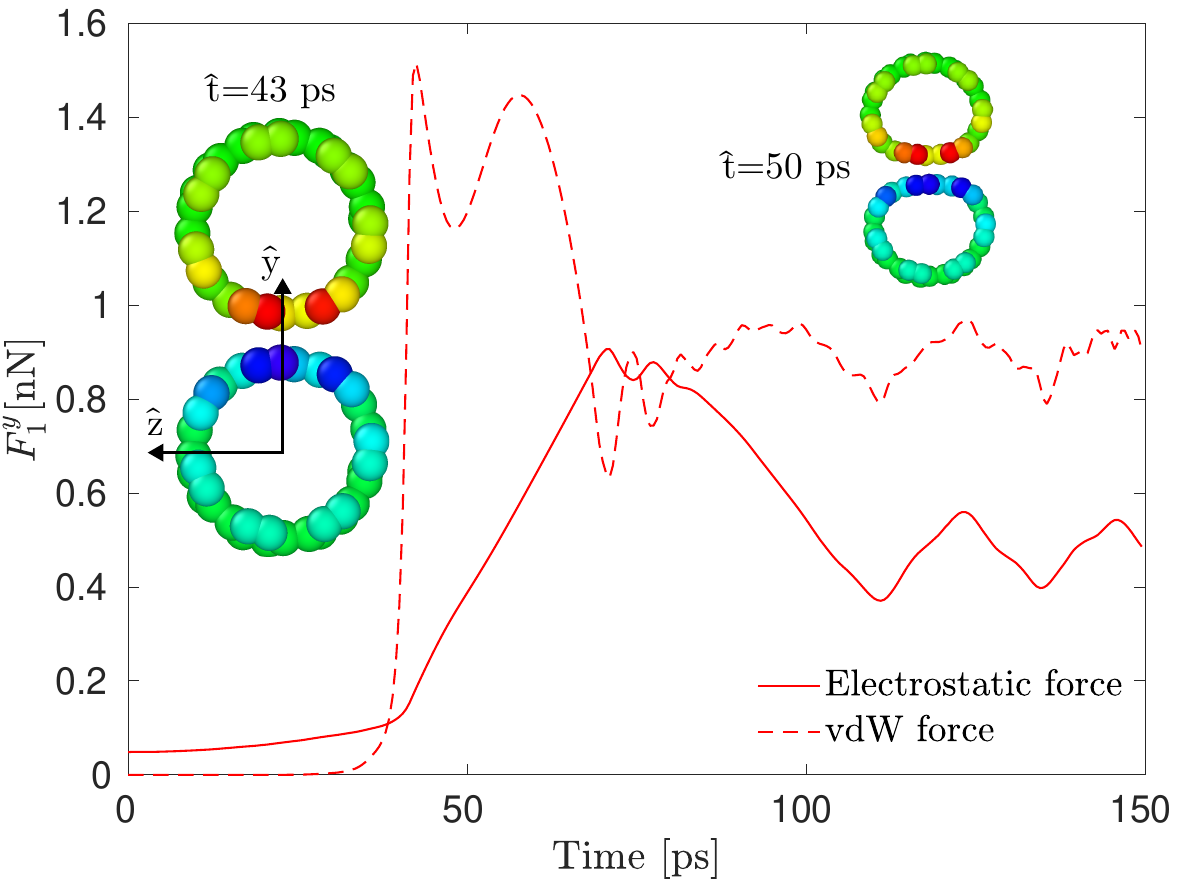}%
}\hfill
\subfloat[]{%
  \includegraphics[width=0.48\columnwidth]{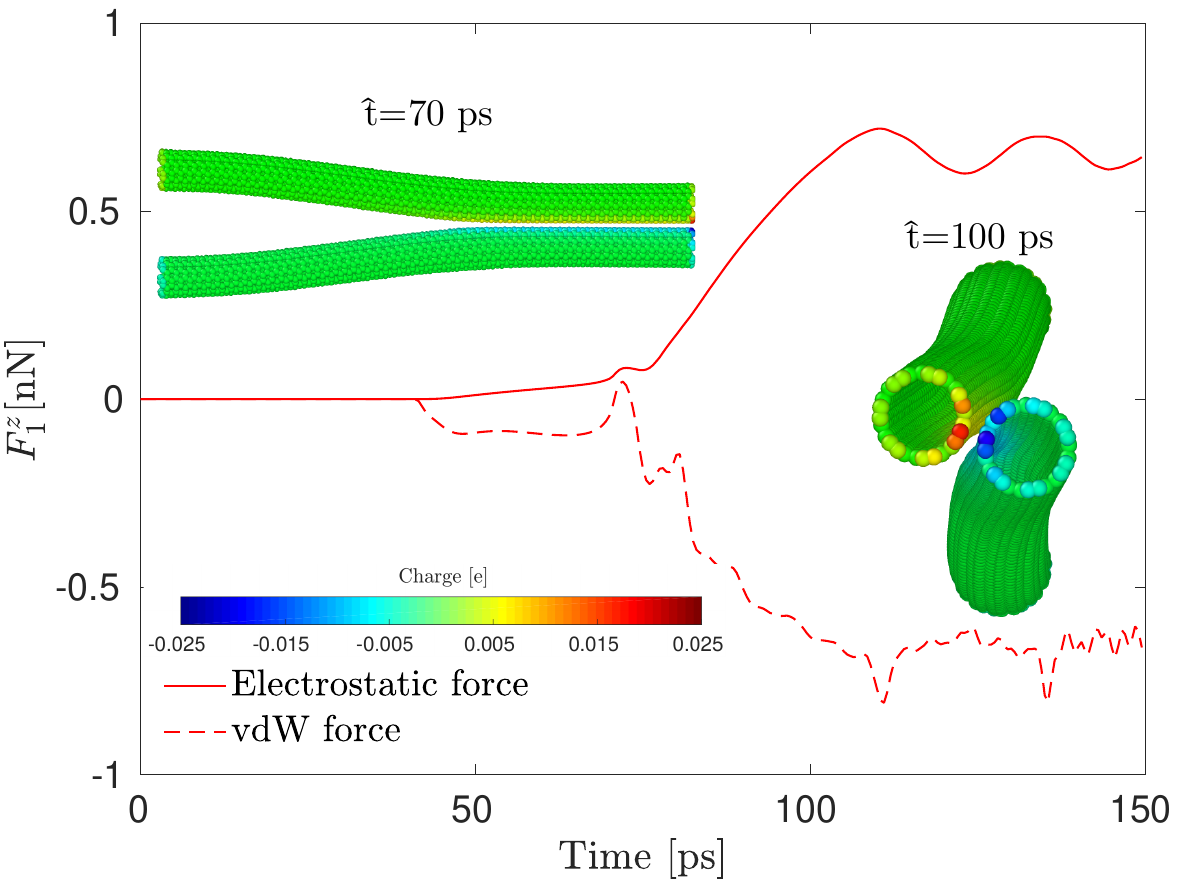}%
}\hfill 
     \caption{Variation of the electrostatic and vdW forces with time acting on SWCNT-1 along the (a) $\hat{y}$- and (b) $\hat{z}$-directions when the applied voltage is equal to the pull-in voltage. The cross-section at the free end is shown in the inset.  \label{f_e_v_C1}}
\end{figure}

Figure~\ref{fixed_free_Contact} shows neutral axis deformations determined from the nonlinear EBBT with end charge concentration employing five, six, and seven modes and centroidal lines of SWCNTs obtained from the MD simulations at the \textit{pull-in voltage}.
To predict the flat configuration of the beam from the free end, utilizing at least seven modes in the beam model is essential. The contact lengths determined from both methods differ by $\sim 20\%$. Figure~\ref{f_e_v_C1} shows the variation of the vdW and electrostatic forces acting on the SWCNT-1 along $\hat{y}$- and $\hat{z}$- directions with time when the applied voltage is equal to the \textit{pull-in voltage}. At this voltage, the two SWCNTs get attracted toward each other over a period of time. At $\hat{t}=0$, the vdW forces are negligible compared to the electrostatic forces. However, these forces rise gradually as the distance between the two SWCNTs decreases. Unlike the fixed-fixed boundary condition, the vdW forces are more dominant than the electrostatic forces at the pull-in. After this point, the highly concentrated forces at the free end lead to deforming both SWCNTs beyond their vdW equilibrium distance. As a result, the vdW repulsive forces start to build up between the atoms of the SWCNTs, which are on the contacting side. 
However, the rest of the atoms of the SWCNTs are subjected to net attractive electrostatic and net attractive vdW forces. Under these distributed forces an SWCNT or beam deflects (downward for SWCNT - 1 and upward for SWCNT-2), however, soon these deflections are resisted by distributed repulsive forces near the contacting side. These repulsive forces resisting the defections in a SWCNT or beam, start flattening SWCNT's deformed geometry from their free ends. We term this action as \textit{zipping} of SWCNTs and it progresses towards the fixed ends. Zipped SWCNTs up to 70 ps are shown in the inset of Fig.~\ref{f_e_v_C1}(b). Further, the cross-sections of the SWCNTs in MD simulations are found deform as shown in the inset of Fig.~\ref{f_e_v_C1}(a), at 50 ps. The circumferential deformations bring more and more atoms of the SWCNTs closer to each other and thus increase the nonbonded forces, this feature cannot be modeled using center-line based EBBT. Hence, the contact lengths determined from the beam model and MD simulations disagree. 

Figure~\ref{f_e_v_C1}(b) shows the variation of transverse force $F_1^z$ acting on SWCNT-1 with time. The figure shows that the transverse force is close to zero until the initiation of the pull-in. After the zipping of the two SWCNTs is completed the transverse wave in $\hat{x}-\hat{y}$ plane continues to travel until it is reflected from the fixed boundary. During this phase, the two SWCNTs oscillate about their new equilibrium configurations in the $\hat{x}-\hat{y}$ plane. As a result, the nonbonded forces fluctuate after 70 ps, as shown in Fig.~\ref{f_e_v_C1}(a). While returning, the transverse wave perturbs the zipped portion, causing the two SWCNTs to slip over each other, as shown in the inset of Fig.~\ref{f_e_v_C1}(b). Slipping of the SWCNTs causes the appearance of transverse force in $\hat{z}$, or the out-of-plane direction as evident from Fig.~\ref{f_e_v_C1}(b).

\begin{figure}[h]
\centering
\subfloat[]{%
  \includegraphics[width=0.49\columnwidth]{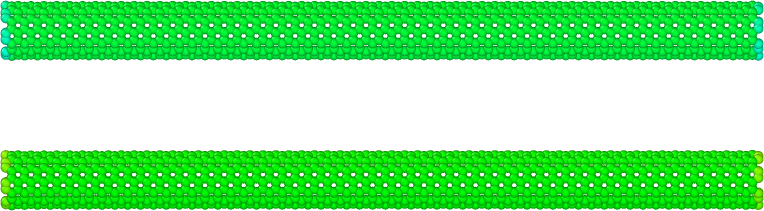}%
}\hfill
\centering
\subfloat[]{%
  \includegraphics[width=0.49\columnwidth]{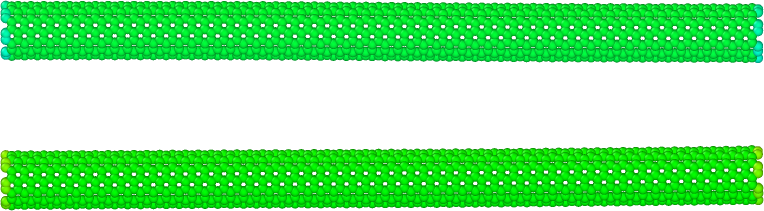}%
}\hfill
\centering
\subfloat[]{%
  \includegraphics[width=0.49\columnwidth]{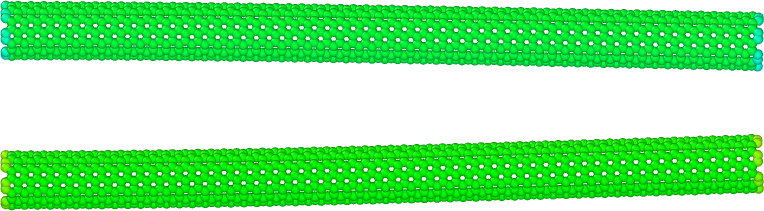}%
}\hfill
\centering
\subfloat[]{%
  \includegraphics[width=0.49\columnwidth]{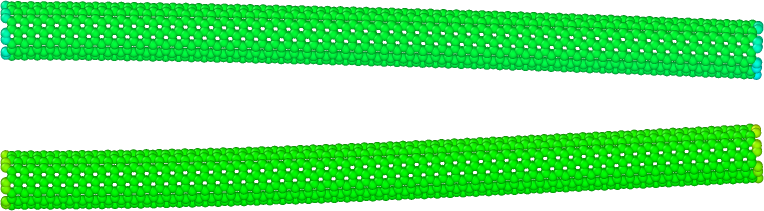}%
}\hfill
\centering
\subfloat[]{%
  \includegraphics[width=0.49\columnwidth]{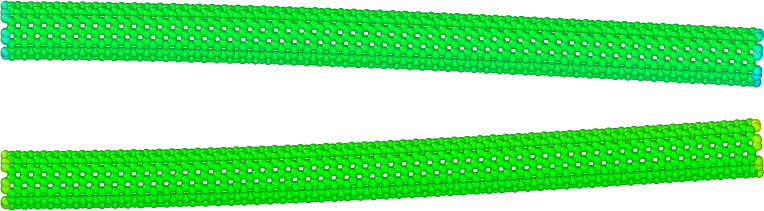}%
}\hfill
\centering
\subfloat[]{%
  \includegraphics[width=0.49\columnwidth]{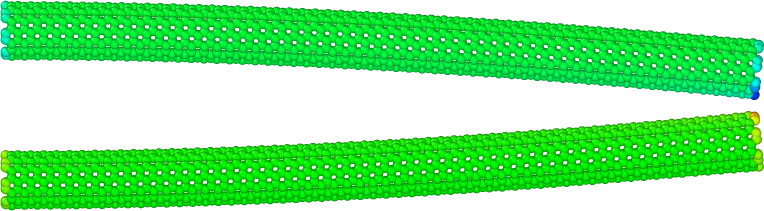}%
}\hfill
\centering
\subfloat{%
  \includegraphics[width=0.7\columnwidth]{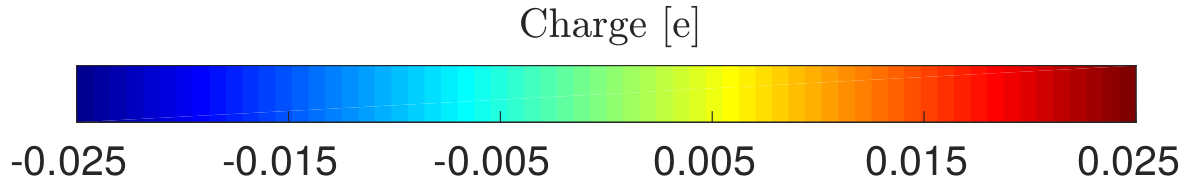}%
}\hfill
        \caption{Variation in configuration and charge contour with time (a) 0 ps (b) 9 ps (c) 18 ps (d) 27 ps (e) 36 ps and (f) 43 ps at \textit{pull-in voltage}.\label{fig:config_cant}}
\end{figure}
Figure~\ref{fig:config_cant} shows the variation of the SWCNTs configuration and charge distribution on them with time when the applied voltage is equal to the \textit{pull-in voltage}. At $t=0$ ps, the charge distribution is uniform when the two SWCNTs are undeformed. However, the charge accumulates significantly at the free end as the distance between the two SWCNTs decreases (see Fig.~\ref{fig:config_cant}). The free end charge is $\sim85 \%$ higher than that at the fixed end at the pull-in. Further, the atoms in contact have $ \sim 270 \%$ more charge than that at the fixed end.

\section{Conclusions} \label{Conclusions}
In this work, the pre- and post-pull-in behaviors of a coupled SWCNT resonator/tweezer employing ReaxFF potential and a geometrically nonlinear EBBT model are studied. The time evolution of charges on the surfaces of SWCNTs are calculated using the QEq scheme within ReaxFF. In the beam model, the ansatz functions for the nonbonded forces are calibrated using MD simulations data. It is found that a nonlinear beam model is essential for predicting the correct value of pull-in voltage for a resonator with significant initial separation gaps. 
The pull-in instabilities in the coupled SWCNTs to the applied voltage are governed by accumulated charge at the mid-span for the resonator with fixed-fixed boundary conditions. In the case of a tweezer with fixed-free boundary conditions, the geometric nonlinearity does not affect the average deflections and pull-in voltages. In this case, it is found that incorporation of the end charge concentration is essential in determining the correct pull-in voltages through the EBBT model. 

The MD simulations also revealed the response of the SWCNTs beyond pull-in for both resonator and tweezer modes of operations. It is found that the circumferential deformations of the SWCNTs become significant beyond pull-in and could lead to the malfunctioning of a resonator or tweezer. Application of excessive voltage in the case of resonators would lead to permanent adhesion/welding of the SWCNTs and in tweezers it will cause slippage of the SWCNTs. 

The post pull-in response presented in this work can be used to develop high-fidelity continuum shell models of SWCNT based resonators/tweezers.

\appendix 
\section{vdW energy}\label{appendix_b-1}
\begin{figure}[!htbp]
\centering
  \includegraphics[width=0.5\columnwidth]{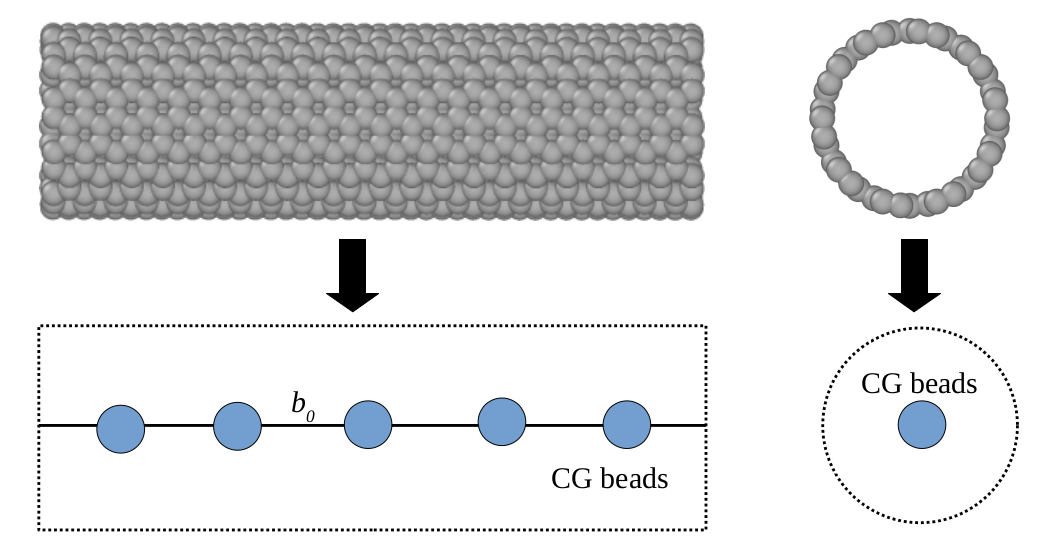}%
\caption{Schematic representation of the CG model.    \label{cg_model}}
\end{figure}

Here, the calibration procedure of coarse-grained (CG) 18-24LJ potential is discussed. In MD simulations, the vdW interactions between carbon atoms of two SWCNTs are modeled using distance-corrected Morse forcefield (see \citet{Chenoweth2008, Strachan2003}). Due to its complex potential form, Coarse-Grained (CG) 18-24LJ potential \citep{ji2019} has been implemented in the beam model. A schematic full-atomic and their CG model representation of SWCNT is shown in Fig.~\ref{cg_model}.
In this model, rings of SWCNT's atoms are considered as a CG bead aligned with the central axis, and each bead is connected to other beads with an equilibrium bond length of $b_0$, as shown in Fig.~\ref{cg_model}. The vdW energy per unit length from the CG model is given by \citep{ji2019}
\begin{equation} \label{cg_ene}
\ds \hat{E}_{\text{vdW}}=\frac{256}{27}\epsilon_{18-24}\hat{\rho}_1^2\left(\frac{88179 \pi\sigma^{24}_{18-24}}{2^{19}\tilde{h}^{23}}-\frac{6435 \pi\sigma^{18}_{18-24}}{2^{15}\tilde{h}^{17}} \right)~,
\end{equation}
where $\tilde{h}$ is the distance between the two central axes of SWCNTs, $\epsilon_{18-24}$ and $\sigma_{18-24}$ are the potential depth and equilibrium distance, respectively. 

\begin{figure}[!htbp]
\centering
\subfloat[]{%
  \includegraphics[width=0.48\columnwidth]{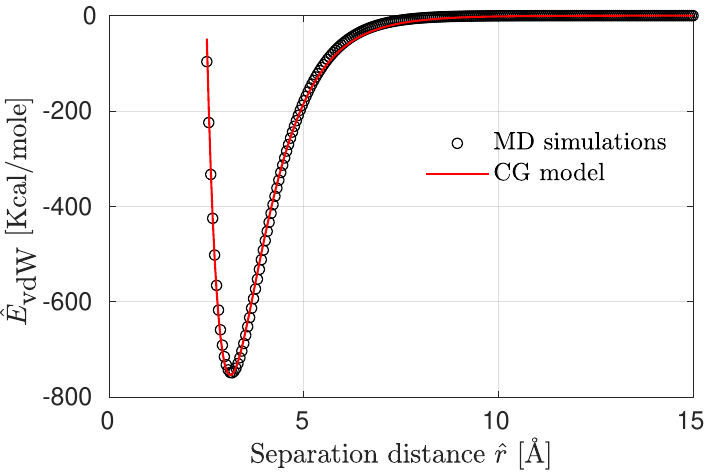}%
}\hfill
\subfloat[]{%
  \includegraphics[width=0.48\columnwidth]{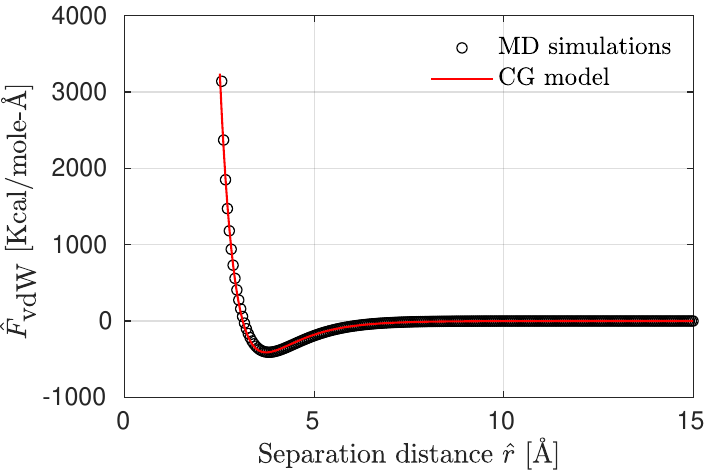}%
}\hfill 
     \caption{Comparison of total (a) vdW energy and (b) vdW force as a function of separation distance determined from MD simulations and the CG model. \label{vdw_cg_ene_force}}
\end{figure}

The two parallel SWCNTs considered in this work are moved rigidly from an initial separation distance of 15 \AA. The vdW energy as a function of surface-surface separation distance $\hat{r}$ is recorded using full atomic MD simulations. The obtained $\hat{E}_\textnormal{vdW}$ vs. $\hat{r}$ data is then fitted into Eq.\eqref{cg_ene} keeping $\epsilon_{18-24}$ and $\sigma_{18-24}$ as variables using least square curve fit in MatLab \citep{matlab}. This fit of total vdW energy is shown in Fig.~\ref{vdw_cg_ene_force}(a). The comparison of vdW force determined from Eq.~\eqref{e:vdW_force} and the MD simulations is given in Fig.~\ref{vdw_cg_ene_force}(b), and it shows good agreement. 

\bibliographystyle{apalike}
\bibliography{Bib/tweezer}

\end{document}